\documentclass[11pt]{article}

\usepackage[final]{acl}

\usepackage{times}
\usepackage{latexsym}

\usepackage[T1]{fontenc}

\usepackage[utf8]{inputenc}

\usepackage{microtype}

\usepackage{inconsolata}

\usepackage{graphicx}

\usepackage{amsmath,amsfonts,bm}
\usepackage{graphicx}
\usepackage{multirow}
\usepackage{subcaption}
\usepackage{wrapfig}
\usepackage{soul}
\usepackage{adjustbox}
\usepackage{enumitem} 
\usepackage{multicol}

\usepackage{array}

\title{CodeTD: Topology of Attention Detects Hallucinations in Code LLMs}

\author{
  \textbf{Daria Voronkova\textsuperscript{1,2}},
  \textbf{Ilya Trofimov\textsuperscript{1}},
  \textbf{Anton Dmitriev\textsuperscript{1}},
  \textbf{Eduard Tulchinskii\textsuperscript{1}},
  \\
  \textbf{Evgeny Burnaev\textsuperscript{1,2}},
  \textbf{Serguei Barannikov\textsuperscript{1,3}}
\\
\\
 \textsuperscript{1}Applied AI Institute
 \textsuperscript{2}AXXX
 \textsuperscript{3}IMJ CNRS
}

\begin{document}
\maketitle
\begin{abstract}
As AI-code assistant tools become widespread, automatic assessment of the correctness of generated code becomes a significant challenge. Code LLMs are prone to hallucinations, which may lead to code that does not solve the required problem, or even to code with severe security vulnerabilities. 
In this paper, we introduce CodeTD -- the first approach to pre-execution assessment of code correctness based on topological data analysis (TDA) of Code LLMs' attention maps.
Our method quantifies prompt-generation mismatch using topological patterns of attention maps. 
We carry out experiments with common benchmarks (HumanEval, MBPP, BigCodeBench, MultiPL-E), 5 programming languages and 10 Code LLMs of size up to 34B parameters.
The experimental results show that the proposed method outperforms recent baselines. Moreover, CodeTD is transferable between coding benchmarks.
\end{abstract}

\section{Introduction}
Code LLMs have revolutionized software engineering, yet they remain prone to hallucinations of various types.
For example, syntactic and runtime errors prevent proper program execution, while logical errors lead to incorrect solutions of the problem.
In some cases, the generated code might contain security issues or robustness issues, such as a memory leak. 
In this paper, we assume that \textit{code hallucination} is a code which is not \textit{functionally correct}, that is, doesn't solve a problem provided in a prompt.  The correctness can be checked by running functional tests. However, in practice functional tests might be (a) unavailable, (b) incomplete, (c) expensive, or (d) unsafe to execute due to security concerns. 
In many applications creating functional tests is a challenging task: user interfaces, distributed/multi-threaded systems, microservice architecture, real-time systems, machine learning/AI systems with stochastic behavior.

For widespread adoption of Code LLMs, there is a strong need for pre-execution verification of code correctness. With current technology, a significant amount of time is spent on debugging and automatic rewriting of generated code \citep{liang2024large}.
We hypothesize that code correctness is intrinsically linked to the structural coherence of the model's attention flow. Unlike scalar metrics (e.g., entropy), topological features capture the connectivity patterns between the problem specification (prompt) and the solution (generation), offering a robust signal for hallucination detection. Attention maps of LLMs have been shown to capture semantically meaningful information and serve as a proxy of the model's ``thinking process''. Previous studies have shown that transformer attention maps are useful for artificial text detection \citep{kushnareva2021artificial}, acceptability judgment \citep{cherniavskii2022acceptability}, and speech classification \citep{tulchinskii2022topological}. 

Only a few methods such as CodeJudge \citep{tong2024codejudge} are specific to detecting hallucinations in Code LLMs, while the research community mostly focuses on general methods for preventing and detecting hallucinations
\citep{peng2023check, zhang2024self, feng2024don, zhang2024truthx, yehuda2024interrogatellm}.

Our contributions are the following:
\begin{itemize}[topsep=0pt,noitemsep,nolistsep, partopsep=0pt, parsep=0ex]
    \item We propose CodeTD (Code Topology Divergence), a new approach to detecting hallucinations in LLM-generated code based on analyzing topology of attention maps;
    \item We carry out computational experiments with 10 Code LLMs of size up to 34B parameters, four benchmarks (HumanEval, MBPP, BigCodeBench, MultiPL-E), 5 programming languages, and show that the proposed method outperforms baselines;
    \item We empirically show that the proposed CodeTD classifier is transferable between code benchmarks and Code LLMs, suggesting robustness against in-distribution overfitting;
    \item We empirically show that features in some attention heads are systematically good indicators of hallucinations across several programming languages.
\end{itemize}

We release our code: \url{https://github.com/VoronkovaDasha/CodeTD}

\section{Related Work}

Code generation via LLMs is a topic of active research. Popular projects include CodeLlama \citep{roziere2023code},
StarCoder2 \citep{lozhkov2024starcoder},
DeepSeek-Coder \citep{guo2024deepseek},
Qwen2.5-Coder \citep{hui2024qwen2}, to name a few. Code LLMs differ by the data used for training, tokenizers, training and fine-tuning protocols (like RLHF), variants of attention mechanism, etc. 

Several works studied attention maps in transformer-based LLMs \citep{clark2019does, htut2019attention, michel2019sixteen}.
\citet{zhang2024eyetrans} proposed to merge human and machine attention for neural code summarization.

The phenomenon of code hallucinations is studied and categorized in several papers.
\citet{tian2024codehalu} introduces a categorization of code hallucinations into four main types: mapping, naming, resource, and logic hallucinations, with each category further divided into different subcategories. \citet{tian2024codehalu} proposed the CodeHalu dataset and studied the frequencies of different types of hallucinations in popular Code LLMs.  
\citet{liu2024exploring}, \citet{jiang2024collu} introduced code hallucination benchmarks.
\citet{liu2024exploring} categorized hallucinations as intent conflicting, inconsistency, repetition, knowledge conflicting, dead code.
\citet{jiang2024collu} found that code LLMs are less confident when hallucinating, since hallucinated tokens have a lower probability and hallucinated generation steps have a higher entropy. \citet{tong2024codejudge} proposed to guide an LLM to work in the ``slow thinking'' regime to obtain a more accurate evaluation of generated code correctness. 

In the broader context of NLP, several works have introduced methods for preventing and detecting hallucinations. \citet{peng2023check} proposed LLM-AUGMENTER, which grounds responses in external knowledge. \citet{zhang2024self} introduced Self-Eval, prompting an LLM to validate its own responses using internal knowledge. \citet{feng2024don} developed hallucination detection through cooperative or competitive model collaboration. \citet{zhang2024truthx} improved truthfulness by editing internal representations during inference. \citet{yehuda2024interrogatellm} proposed InterrogateLLM, which detects inconsistency by reconstructing queries from generated answers. \citet{bazarova2025hallucination} introduced TOHA, which quantifies structural properties of attention graphs to detect hallucinations in Retrieval-Augmented Generation (RAG). We provide a detailed comparison with \citep{bazarova2025hallucination} in Appendix \ref{app:bazarova}.

\section{Code Hallucination as Functional Incorrectness}
 
In this paper, we assume that \textit{code hallucination} is a code which is not \textit{functionally correct}, that is, does not solve the problem provided in a prompt.
The correctness can be checked by running functional tests. However, in practice functional tests might be (a) unavailable, (b) incomplete, (c) expensive, or (d) unsafe to execute due to security concerns. 
The goal of our work is to develop a pre-execution verifier of code correctness. 
Hallucination types can be roughly classified into syntactic, runtime and logical errors.
For modern Code LLMs, the major problem is logical errors (the output does not transform the input as specified in the prompt) which correspond to $\sim$78\% of incorrect code cases (see Appendix \ref{app:error_types}). \citet{tian2024codehalu} do not include syntax errors into code hallucinations. However, modern Code LLMs generate syntactically incorrect code in rare cases. Thus, our definition almost coincides with \cite{tian2024codehalu}. 
So, static code analysis tools are of little help. Other definitions of code hallucinations exist \citep{jiang2024collu, liu2024exploring}, but our work is focused on what matters in practice -- detecting the outcome (incorrectness) regardless of the root cause.

\section{Background. Transformer-based LLMs}

All state-of-the-art Code LLMs are based on different variants of the transformer architecture \citep{vaswani2017attention}. 
A transformer architecture comprises $L$ layers of multi-head self-attention blocks, each of them having $H$ heads. Each attention head takes the matrix $X \in \mathbb{R}^{n\times d}$ as an input, and an output is $X^{out} = A (X W^{V})$, where
\begin{align*}
A = \text{softmax}\left( \frac{(X W^Q) (X W^{K})^T}{\sqrt{d}} \right),
\end{align*}
and $W^Q, W^K, W^V \in \mathbb{R}^{d\times d}$ are projection matrices, and $A \in \left[0,1 \right]^{n\times n}$ is an \textbf{attention map}. 
In the self-attention block, the attention map shows how each token in the input sequence ``interacts'' with every other token in the same sequence.
A token might attend more to other tokens that are contextually related. We interpret each element $a_{i,j}$ of an attention map as an ``interaction force'' between tokens $i$ and $j$.

\section{Motivation}

\begin{figure}[t]
\centering
  \includegraphics[width=\columnwidth]{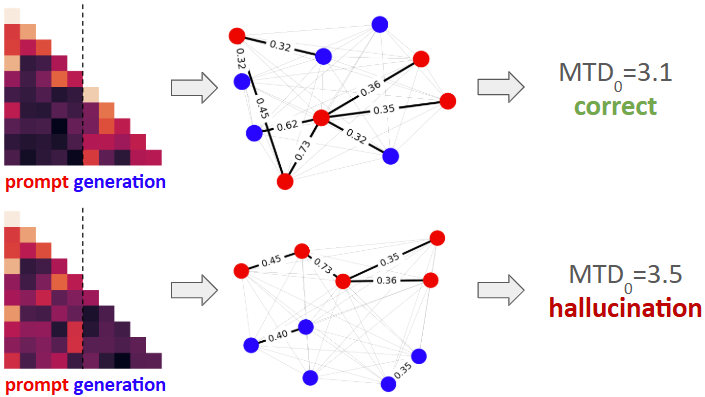}
  \caption{Two attention maps (left) have different prompt-generation connectivity strength. Attention maps are presented as weighted graphs (middle) where weights equal to attention values. Tokens of prompt and generation are depicted in red and blue respectively. Edges with weights $\ge0.3$ are shown with a bold line. Graph layout places vertices with high connecting weight closer. Correct code has high prompt-generation connectivity (top), while hallucinated code doesn't (bottom). MTD$_0$ is an integral characteristic of prompt-generation connectivity.}
  \label{fig:motivation}
  \vskip-2pt
\end{figure}

\label{sec:motivation}
\textbf{Attention Map as a Weighted Graph}.
While an attention map is typically represented as a matrix, we treat it as a weighted graph. For $n$ tokens in a sequence, we consider a fully-connected weighted graph with $n$ vertices, where edge weights are related to the ``interaction force'' between tokens (vertices).

\textbf{Code Hallucinations.}
In the context of code generation, we naturally have two sets of tokens: a prompt and a generation.
A common cause of hallucinations is when the model’s attention drifts away from the prompt.
We hypothesize that it means low  connectivity between prompt-generation tokens.
Figure \ref{fig:motivation} illustrates our intuition. Attention maps with high/low prompt-generation connectivity are shown in Figure \ref{fig:motivation} top/bottom respectively.
A straightforward baseline would threshold attention scores to retain only strongly interacting tokens. However, this approach requires an unknown optimal threshold and introduces discontinuous topological changes with the change of a threshold or weights.

\begin{figure*}[th!]
\begin{center}
  \includegraphics[width=\textwidth]{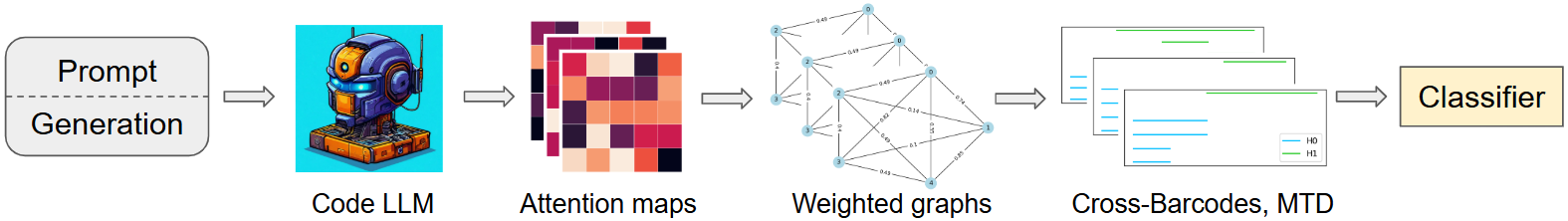}
  \caption{A pipeline of the proposed CodeTD method for hallucination detection: (1) a prompt concatenated with a generated code is fed into a Code LLM. (2) Attention maps from the Code LLM are obtained. (3) Attention maps are transformed into fully-connected weighted graphs. (4) Cross-Barcodes and MTD features for weighted graphs are calculated. (5) On the top of the generated features a binary classifier of hallucinations is fitted.}
  \label{fig:pipeline}
\end{center}
\end{figure*}

\begin{figure*}[t]
\begin{center}
\begin{subfigure}[t]{0.49\textwidth}
\includegraphics[width=\columnwidth]{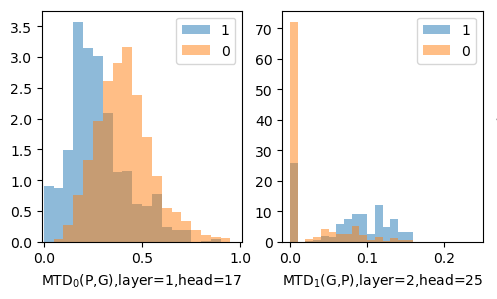}
  \caption{HumanEval}
\end{subfigure}
\begin{subfigure}[t]{0.49\textwidth}
\includegraphics[width=\columnwidth]{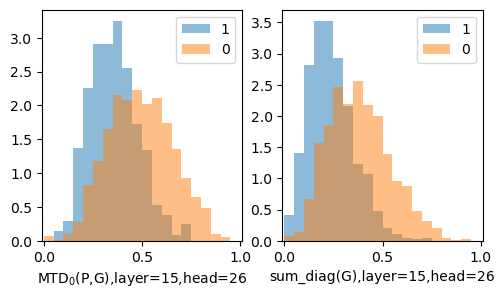}
  \caption{MBPP}
\end{subfigure}
\caption{Distribution of classes (0-code is not correct, hallucination; 1-code is correct) vs. features from attention maps. Some of the most discriminative features are presented. Features are normalized with MinMaxScaler. Features are extracted from CodeLlama-7B.}
\label{fig:attn_features_codellama}
\end{center}
\end{figure*}

\section{Method}
\label{sec:methods}

\textbf{Manifold Topology Divergence}. To measure prompt-generation connectivity we use
MTD (Manifold Topology Divergence) \citep{Barannikov2021manifold}. MTD is a tool of Topological Data Analysis \citep{Chazal2019introduction} that can be used to evaluate the ``dissimilarity'' between two disjoint sets of vertices $V = P \sqcup G$ in a weighted graph $\mathcal{G} = (V, E, W)$ or, in other words, to which degree one set of vertices is connected to another set. 
MTD is computed via the following algorithm: (1) take vertices of $P \sqcup G$ with edges connecting vertices of $P$ (2) add the rest of edges in ascending order by their weights (3) track changes of topological patterns (connected components, cycles\footnote{Appendix \ref{app:cycles} explains why cycles matter.}, etc.) which appear and disappear (are ``born'' and ``die'').
The output of this algorithm is a Cross-Barcode$_k$ -- a multi-set of  birth-death moments. Its integral characteristic is MTD$_k$ -- a sum of birth-death intervals' lengths (step-by-step calculation are shown in Fig. \ref{fig:mtd_evolution}, \ref{fig:mtd_barcode})
$$
\text{MTD}_k = \sum_{(b_i, d_i) \in CrossBarcode_k} d_i - b_i.
$$
Here index $k$ refers to a dimensionality of topological patterns: 0-connected components, 1-cycles, etc. The higher MTD$_k$ is, the greater is the ``dissimilarity'' between sets of tokens.
MTD$_k$ scores of some heads have a significant discriminative power and are shown in Figure~\ref{fig:attn_features_codellama}.
MTD$_k$, as a kind of persistence barcode, enjoys stability w.r.t. perturbations of weights \citep{cohen2005stability}.
We refer the reader to Appendix \ref{app:mtd} for precise definitions. 

\textbf{Algorithm.}
Specifically, for $n$ tokens in a prompt+generation sequence, we consider a fully-connected undirected weighted graph with $n$ vertices.
We define the edge weights as $w_{i,j} = 1 - a_{i,j}$ for $i > j$ (we use decoder-only LLMs with causal attention). Since the MTD filtration process adds edges in ascending order of weight, this transformation ensures that tokens with high attention scores (low $w_{i,j}$) are connected early in the filtration process. This allows the topological summary to prioritize strong semantic interactions between prompt and generation tokens.
Then, Cross-Barcode$_k$ and MTD$_k$ for a weighted ``attention graph'' can be calculated\footnote{Appendix \ref{app:barcodes} shows examples of Cross-Barcodes and corresponding attention maps.}.
Weak prompt-to-generation attention indicates that the generated tokens are not semantically grounded in the prompt.
MTD measures the ``semantic coupling'' between the specification (prompt) and the implementation (generation). A low MTD score indicates the generation is structurally detached from the requirements.
That is, Code LLM drifts away from the prompt during generation and hallucinates.

To predict code hallucinations, we use the following set of features which are calculated for every layer and head of a Code LLM:
\begin{itemize}
    \item $\text{MTD}_0(P, G) / |G|$, $\text{MTD}_0(G, P) / |P|$
    \item$\text{MTD}_1(P, G) / |G|$, $\text{MTD}_1(G, P) / |P|$
    \item $\sum_{i\in P} a_{i,i} / |P|$, $\sum_{i\in G} a_{i,i} / |G|$
\end{itemize}
To ensure scale invariance across varying prompt and generation lengths, all topological features are normalized by the cardinality of the corresponding vertex sets ($|P|$ and $|G|$).
In addition, averages of diagonal values of the attention matrices that are not directly present in edge weights are included.  At the top of the proposed topological features, we trained an XGBoost \citep{chen2016xgboost} classifier\footnote{See Appendix \ref{app:clf_ablation} for a classifier ablation study.}. Figure \ref{fig:pipeline} presents the high-level pipeline of CodeTD.
CodeTD is a supervised method requiring labels derived from functional tests for training; crucially, no test execution is required at inference, and the classifier is transferable across benchmarks.

\section{Experiments}
\subsection{Dataset Construction}
\label{sec:gen_data}

\textbf{Code LLMs and benchmarks.}
In our main experiments, we use the following popular code LLMs: StarCoder2-7B \citep{lozhkov2024starcoder}, CodeLlama-7B \citep{roziere2023code},
DeepSeek-Coder-6.7B \citep{guo2024deepseek},
Qwen2.5-Coder-7B \citep{hui2024qwen2}, Magicoder-S-DS-6.7B \citep{wei2024magicoder}. Additionally, we verify the proposed approach on smaller (Qwen2.5-Coder-1.5B, Qwen2.5-Coder-3B) and larger (CodeLlama-34B, DeepSeek-Coder-33B, Qwen2.5-Coder-32B) models.
We adapted public benchmarks for evaluation of code generation: HumanEval (HE) \citep{chen2021evaluating}, MBPP \citep{austin2021program}, BigCodeBench (BCB) \citep{zhuo2024bigcodebench}, MultiPL-E \citep{cassano2023multiple}.

\textbf{Generation procedure.}
To provide diversity and increase dataset size we do temperature sampling with $T = 0.8$ for each of the coding problems: 
for the small-sized (1.5B-3B) and medium-sized (6.7B-7B) models, we obtained 25 generations per task for HumanEval, 5 generations per task for MBPP, 1 generation per task for BCB, and 10 generations per task for MultiPL-E; for larger models (32B-33B), we used 10 generations per task for HumanEval. To address the quality of the proposed approach in different prompting regimes, we used a 0-shot prompt for the HumanEval dataset and a 1-shot/2-shot prompt for the MBPP dataset\footnote{This follows a standard experimental setting of these benchmarks: 0-shot for HumanEval and few-shot for MBPP, e.g. \cite{roziere2023code}, \cite{guo2024deepseek}.}.
In addition, we provide evaluation with \textbf{greedy decoding} to account for a more practical setup, see Appendix \ref{sec:greedy}. CodeTD outperforms baselines in this setting.
See Appendix~\ref{app:gen_procedure} for further details on generation procedure.

\textbf{Dataset statistics.}
Table \ref{tab:gen_data_stat_total} presents a summary of code snippets generated.
The correctness of the code is evaluated via functional tests provided together with the coding benchmarks.
Functional tests check that the function called with certain arguments has the corresponding output (examples are shown in Figures \ref{fig:he_gen}, \ref{fig:mbpp_gen} in Appendix~\ref{app:gen_procedure}).
Incorrect code is considered a ``hallucination''; prediction of code's correctness is a binary classification problem. Before moving further, note that there is a strong negative dependency between prompt and generation lengths and code quality, see Fig.~\ref{fig:promt_gen_vs_correctness},~\ref{fig:promt_gen_vs_correctness_mbpp}.
The longer the prompt (i.e. task description) and generation (i.e. task solution) are, the lower is the probability of code's correctness. 
These attributes are natural baselines for hallucination's prediction.

\subsection{Code Hallucination Detection}
\label{sec:clf_quality}

\textbf{Train/test pipeline.}
Using the generated data, we estimate the classification quality of the proposed approach.
We applied 5-fold stratified group cross-validation where different solutions of the same coding problem belonged to the same group. In this way, training and testing were performed always at non-overlapping coding problems (prompts). The reported results are the mean and std. deviation estimated over the $5$ folds. 

\textbf{Baselines.}
We used the following baselines for comparison:

\textbf{supervised:}
\begin{itemize}[topsep=0pt,noitemsep,nolistsep, partopsep=0pt, parsep=0ex]
    \item XGBoost classifier trained on tokenized prompt length, tokenized generation length;
    \item a linear classifier on top of a frozen CodeT5-base encoder \cite{wang2021codet5identifierawareunifiedpretrained};
    \item attention-based hallucination detectors: AttnLogDet, AttnEigvals, LapEigvals \citep{binkowski2025hallucination}.
\end{itemize}

\textbf{unsupervised/zero-shot:}
\begin{itemize}[topsep=0pt,noitemsep,nolistsep, partopsep=0pt, parsep=0ex]
    \item mean log. probability of generated tokens \cite{chen2021evaluating};
    \item Pylint\footnote{https://www.pylint.org/}, a static code analysis tool for Python;
    \item Self-Eval \cite{zhang2024self};
    \item CodeJudge \cite{tong2024codejudge}.
\end{itemize}

See Appendix~\ref{app:train_details} for  further training details.

\textbf{Discussion.}
Table~\ref{tab:detection_results_temperature} presents our main results. 
In the majority of cases, the proposed CodeTD classifier based on the features of the attention maps performed significantly better than the baselines and demonstrated stable results for all models and datasets as measured by the ROC-AUC score. 
Thus, we believe that the proposed topology-based attention features are strong enough to capture the most important information for code hallucination detection. 
Some individual features made a significant contribution to classification quality, see Figure~ \ref{fig:attn_features_codellama}. See Appendix \ref{app:complexity} for details on computational complexity.
CodeTD outperforms baselines on \textbf{BigCodeBench (BCB)}, the most practically relevant benchmark. 
The improvement over the runner-up (LapEigvals) is significant via Fisher's combined test with $p = 0.017$ (See Appendix \ref{app:bcb_stat_significance} for full data).
BCB is a stress-test for CodeTD since the proportion of correct generations can be as low as 5.9\% (Table \ref{tab:gen_data_stat_total}).

\textbf{Experiments on the MultiPL-E benchmark.}
We evaluate CodeTD across languages with varying resource availability and ecosystem maturity: Java (high-resource), Go (medium), Rust (low), and Lua (niche) of the HumanEval subdivision of the MultiPL-E dataset \citep{cassano2023multiple}. 
We provide comparison with attention-based hallucination detectors as the strongest baselines in Table \ref{tab:more_languages_detect_rocauc_mean}. The proposed CodeTD classifier detects hallucinations even in low resource and niche languages, and it outperforms other attention-based methods in approximately $84\%$ of cases. Furthermore, there are several single features that exhibit a high predictive performance consistently across all programming languages, see Appendix \ref{app:more_languages}. 

\textbf{Evaluation with larger Code LLMs.} The proposed CodeTD consistently outperforms CodeJudge \cite{tong2024codejudge} with CodeLlama-34B as evaluation model both in greedy decoding and sampling setups (see Table \ref{tab:codejudge}).
Furthermore, the CodeTD outperforms zero-shot prompt evaluation for Qwen2.5-Coder-32B (Table \ref{tab:detection_qwencoder32b}), and proposed topological features of DeepSeek-Coder-6.7B improve the zero-shot prompt evaluation for DeepSeek-R1-671B (Table \ref{tab:detection_deepseek_r1}). See Appendix \ref{app:large_models} for further details.

\begin{table}
\setlength{\tabcolsep}{2pt}
\centering
\begin{adjustbox}{width=0.98\columnwidth}
\begin{tabular}{lccc}
   \hline
   \textbf{Method} & \textbf{HE} & \textbf{MBPP} & \textbf{BCB} \\ \hline
   \multicolumn{4}{c}{DeepSeek-Coder-6.7B} \\ \hline
   Prompt. Len. & $58.0 \pm 6.7$ & $49.5 \pm 4.1$ & $54.9 \pm 2.3$ \\
   Gen. Len. & $58.5 \pm 3.7$ & $55.4 \pm 1.7$ & $54.7 \pm 2.7$ \\
   Mean Log. Prob. & $68.9 \pm 3.0$ & $60.7 \pm 2.2$ & $50.1 \pm 2.9$ \\
   Pylint & $54.7 \pm 1.1$ & $52.8 \pm 0.9$ & $61.2 \pm 1.1$ \\
   CodeT5-base ft. & $63.1 \pm 4.7$ & $52.8 \pm 3.2$ & $54.5 \pm 2.8$ \\
   Self-Eval & $57.3 \pm 4.5$ & $51.0 \pm 0.8$ & $52.9 \pm 1.6$ \\
   AttnLogDet & $78.8 \pm 3.6$ & $77.3 \pm 5.1$ & $67.5 \pm 4.8$ \\
   AttnEigvals & $75.1 \pm 3.2$ & $75.9 \pm 3.5$ & $\underline{67.6 \pm 1.9}$ \\
   LapEigvals & $\underline{81.7 \pm 1.9}$ & $\underline{77.6 \pm 4.5}$ & $64.6 \pm 0.5$ \\
   CodeTD (ours) & $\mathbf{86.4 \pm 1.8}$ & $\mathbf{81.2 \pm 3.7}$ & $\mathbf{69.2 \pm 2.8}$ \\ \hline
      \multicolumn{4}{c}{StarCoder2-7B} \\ \hline
   Prompt. Len. & $56.3 \pm 5.9$ & $55.2 \pm 4.0$ & $51.3 \pm 6.5$ \\
   Gen. Len. & $58.9 \pm 2.4$ & $59.5 \pm 1.2$ & $61.7 \pm 3.1$ \\
   Mean Log. Prob. & $69.4 \pm 1.7$ & $61.9 \pm 3.6$ & $79.4 \pm 3.3$ \\
   Pylint & $58.9 \pm 1.5$ & $53.0 \pm 0.9$ & $59.5 \pm 1.9$ \\
   CodeT5-base ft. & $62.2 \pm 1.8$ & $56.4 \pm 1.9$ & $62.5 \pm 5.3$ \\
   Self-Eval & $50.6 \pm 1.7$ & $59.0 \pm 2.1$ & $46.2 \pm 3.6$ \\
   AttnLogDet & $80.4 \pm 4.3$ & $75.3 \pm 4.2$ & $75.3 \pm 4.1$ \\
   AttnEigvals & $78.2 \pm 2.1$ & $73.2 \pm 3.1$ & $73.8 \pm 2.3$ \\
   LapEigvals & $\underline{80.6 \pm 2.9}$ & $\underline{79.3 \pm 2.8}$ & $\mathbf{85.3 \pm 2.4}$ \\
   CodeTD (ours) & $\mathbf{82.5 \pm 2.1}$ & $\mathbf{81.4 \pm 3.6}$ & $\underline{83.7 \pm 4.3}$ \\ \hline
   \multicolumn{4}{c}{CodeLlama-7B} \\ \hline
   Prompt. Len. & $61.6 \pm 4.4$ & $59.1 \pm 4.2$ & $51.3 \pm 4.3$ \\
   Gen. Len. & $60.1 \pm 5.3$ & $60.8 \pm 2.5$ & $58.7 \pm 2.4$ \\
   Mean Log. Prob. & $64.1 \pm 2.0$ & $61.0 \pm 3.7$ & $57.1 \pm 3.5$ \\
   Pylint & $55.1 \pm 0.3$ & $53.1 \pm 0.5$ & $60.1 \pm 0.8$ \\
   CodeT5-base ft. & $74.5 \pm 6.3$ & $61.7 \pm 3.0$ & $50.7 \pm 3.9$ \\
   Self-Eval & $49.7 \pm 1.7$ & $50.0 \pm 0.0$ & $51.1 \pm 1.8$ \\
   AttnLogDet & $77.3 \pm 5.2$ & $75.6 \pm 1.8$ & $71.4 \pm 2.9$ \\
   AttnEigvals & $80.2 \pm 6.0$ & $75.6 \pm 3.1$ & $68.8 \pm 3.4$ \\
   LapEigvals & $\underline{83.3 \pm 3.6}$ & $\underline{79.7 \pm 2.7}$ & $\underline{73.1 \pm 2.9}$ \\
   CodeTD (ours) & $\mathbf{85.6 \pm 3.9}$ & $\mathbf{83.4 \pm 3.3}$ & $\mathbf{75.0 \pm 3.4}$ \\ \hline
   \multicolumn{4}{c}{Qwen2.5-Coder-7B} \\ \hline
   Prompt. Len. & $56.6 \pm 4.1$ & $53.1 \pm 2.9$ & $51.2 \pm 1.8$ \\
   Gen. Len. & $57.3 \pm 1.1$ & $57.2 \pm 2.5$ & $55.7 \pm 1.8$ \\
   Mean Log. Prob. & $61.4 \pm 3.6$ & $61.6 \pm 0.7$ & $51.5 \pm 3.0$ \\
   Pylint & $63.7 \pm 2.4$ & $62.2 \pm 2.3$ & $63.8 \pm 1.0$ \\
   CodeT5-base ft. & $62.5 \pm 4.8$ & $54.1 \pm 3.1$ & $54.2 \pm 6.1$ \\
   Self-Eval & $73.3 \pm 5.0$ & $66.7 \pm 1.7$ & $52.5 \pm 3.0$ \\
   AttnLogDet & $72.5 \pm 3.6$ & $70.6 \pm 3.8$ & $\mathbf{70.3 \pm 3.3}$ \\
   AttnEigvals & $74.2 \pm 3.5$ & $72.2 \pm 4.3$ & $67.8 \pm 3.5$ \\
   LapEigvals & $\underline{78.4 \pm 2.6}$ & $\underline{76.8 \pm 4.3}$ & $66.4 \pm 0.9$ \\
   CodeTD (ours) & $\mathbf{81.6 \pm 4.7}$ & $\mathbf{80.9 \pm 3.5}$ & $\underline{70.0 \pm 1.4}$ \\ \hline
   \multicolumn{4}{c}{Magicoder-S-DS-6.7B} \\ \hline
   Prompt. Len. & $55.7 \pm 6.7$ & $52.2 \pm 2.0$ & $56.5 \pm 4.9$ \\
   Gen. Len. & $52.2 \pm 3.6$ & $53.7 \pm 1.3$ & $53.9 \pm 4.2$ \\
   Mean Log. Prob. & $68.8 \pm 4.1$ & $60.0 \pm 2.4$ & $54.7 \pm 5.0$ \\
   Pylint & $52.7 \pm 0.9$ & $52.0 \pm 1.2$ & $60.5 \pm 2.0$ \\
   CodeT5-base ft. & $56.8 \pm 4.0$ & $52.3 \pm 3.9$ & $56.5 \pm 5.5$ \\
   Self-Eval & $46.1 \pm 5.1$ & $49.2 \pm 1.1$ & $51.7 \pm 2.7$ \\
   AttnLogDet & $75.8 \pm 2.6$ & $78.3 \pm 3.1$ & $\underline{68.2 \pm 2.6}$ \\
   AttnEigvals & $71.5 \pm 3.4$ & $75.1 \pm 2.7$ & $\mathbf{68.4 \pm 3.2}$ \\
   LapEigvals & $\underline{76.7 \pm 3.9}$ & $\underline{79.5 \pm 1.4}$ & $66.8 \pm 1.0$ \\
   CodeTD (ours) & $\mathbf{79.3 \pm 3.5}$ & $\mathbf{80.7 \pm 2.5}$ & $\mathbf{68.4 \pm 1.5}$ \\ \hline
\end{tabular}
\end{adjustbox}
\vskip-4pt
\caption{ROC-AUC of code hallucination detection for generation with sampling temperature $T = 0.8$.}
\label{tab:detection_results_temperature}
\end{table}

\begin{table}[htb!]
\setlength{\tabcolsep}{2pt}
\centering
\begin{tabular}{lcccc}
   \hline
   \textbf{Method} & \textbf{Java} & \textbf{Go} & \textbf{Rust} & \textbf{Lua}  \\ \hline
   \multicolumn{5}{c}{StarCoder2-7B} \\ \hline
   AttnLogDet & $80.7$ & $82.4$ & $\underline{77.8}$ & $74.6$ \\
   AttnEigvals & $\underline{81.6}$ & $80.4$ & $75.2$ & $77.8$ \\
   LapEigvals & $\underline{81.6}$ & $\underline{82.5}$ & $77.1$ & $\underline{78.8}$ \\
   CodeTD (ours) & $\mathbf{82.5}$ & $\mathbf{86.6}$ & $\mathbf{82.5}$ & $\mathbf{82.0}$ \\ \hline
   \multicolumn{5}{c}{CodeLlama-7B} \\ \hline
   AttnLogDet & $\mathbf{85.2}$ & $55.5$ & $63.4$ & -- \\
   AttnEigvals & $\underline{78.2}$ & $66.7$ & $\underline{76.7}$ & -- \\
   LapEigvals & $75.7$ & $\underline{68.6}$ & $74.3$ & -- \\
   CodeTD (ours) & $76.8$ & $\mathbf{81.9}$ & $\mathbf{77.5}$ & -- \\ \hline
   \multicolumn{5}{c}{DeepSeek-Coder-6.7B} \\ \hline
   AttnLogDet & $80.1$ & $81.9$ & $78.3$ & $74.7$ \\
   AttnEigvals & $78.2$ & $79.0$ & $74.5$ & $84.6$ \\
   LapEigvals & $\underline{83.1}$ & $\mathbf{89.0}$ & $\underline{82.1}$ & $\underline{86.1}$ \\
   CodeTD (ours) & $\mathbf{84.5}$ & $\underline{85.2}$ & $\mathbf{82.3}$ & $\mathbf{86.2}$ \\ \hline
   \multicolumn{5}{c}{Qwen2.5-Coder-7B} \\ \hline
   AttnLogDet & $65.7$ & $76.7$ & $72.4$ & $80.3$ \\
   AttnEigvals & $78.3$ & $85.9$ & $\underline{84.0}$ & $85.8$ \\
   LapEigvals & $\underline{80.8}$ & $\underline{87.9}$ & $79.3$ & $\underline{87.3}$ \\
   CodeTD (ours) & $\mathbf{81.7}$ & $\mathbf{91.1}$ & $\mathbf{87.5}$ & $\mathbf{90.1}$ \\ \hline
   \multicolumn{5}{c}{Magicoder-S-DS-6.7B} \\ \hline
   AttnLogDet & $74.5$ & $70.2$ & $69.8$ & $68.8$ \\
   AttnEigvals & $69.2$ & $70.9$ & $\underline{69.9}$ & $75.4$ \\
   LapEigvals & $\underline{76.2}$ & $\mathbf{84.0}$ & $\mathbf{74.8}$ & $\underline{79.6}$ \\
   CodeTD (ours) & $\mathbf{77.8}$ & $\underline{80.7}$ & $\mathbf{74.8}$ & $\mathbf{80.1}$ \\ \hline
\end{tabular}
\caption{Average ROC-AUC of the proposed method over 5 folds on different programming languages from MultiPL-E dataset. Due to space limitations, see Table \ref{tab:more_languages_detect_rocauc_full} in Appendix for full statistics with std. deviations.}
\label{tab:more_languages_detect_rocauc_mean}
\vskip-10pt
\end{table}

\begin{table}[htb!]
\setlength{\tabcolsep}{2pt}
\centering
\begin{tabular}{lcc}
    \hline
    \textbf{Method} & \textbf{HE $\to$ MBPP} & \textbf{MBPP $\to$ HE} \\ \hline
    \multicolumn{3}{c}{StarCoder2-7B} \\ \hline
   Mean Log. Prob. & $63.8$ & $\mathbf{71.8}$ \\
   CodeT5-base ft. & $53.7$ & $59.1$ \\
   AttnLogDet & $57.7$ & $59.3$ \\
   AttnEigvals & $53.0$ & $60.8$ \\
   LapEigvals & $\underline{65.2}$ & $\underline{66.6}$ \\
   CodeTD (ours) & $\mathbf{67.7}$ & $66.0$ \\
   \hline
    \multicolumn{3}{c}{CodeLlama-7B} \\ \hline
   Mean Log. Prob. & $57.7$ & $65.0$ \\
   CodeT5-base ft. & $54.9$ & $62.4$ \\
   AttnLogDet & $60.2$ & $\underline{68.7}$ \\
   AttnEigvals & $57.8$ & $66.4$ \\
   LapEigvals & $\underline{60.7}$ & $64.2$ \\
   CodeTD (ours) & $\mathbf{69.5}$ & $\bf{80.3}$ \\ 
   \hline
    \multicolumn{3}{c}{DeepSeek-Coder-6.7B} \\ \hline
   Mean Log. Prob. & $62.7$ & $\underline{69.1}$ \\
   CodeT5-base ft. & $53.4$ & $55.9$ \\
   AttnLogDet & $59.9$ & $58.9$ \\
   AttnEigvals & $55.1$ & $64.3$ \\
   LapEigvals & $\underline{63.5}$ & $66.4$ \\
   CodeTD (ours) & $\mathbf{69.9}$ & $\mathbf{72.4}$ \\
   \hline
    \multicolumn{3}{c}{Qwen2.5-Coder-7B} \\ \hline
   Mean Log. Prob. & $60.3$ & $\underline{64.7}$ \\
   CodeT5-base ft. & $49.1$ & $51.6$ \\
   AttnLogDet & $55.8$ & $56.5$ \\
   AttnEigvals & $49.9$ & $57.9$ \\
   LapEigvals & $\underline{65.5}$ & $62.1$ \\
   CodeTD (ours) & $\bf{70.9}$ & $\mathbf{65.8}$ \\
   \hline
    \multicolumn{3}{c}{Magicoder-S-DS-6.7B} \\ \hline
   Mean Log. Prob. & $63.8$ & $\mathbf{69.8}$ \\
   CodeT5-base ft. & $49.3$ & $45.9$ \\
   AttnLogDet & $65.4$ & $65.0$ \\
   AttnEigvals & $59.5$ & $58.6$ \\
   LapEigvals & $\underline{68.1}$ & $\underline{65.9}$ \\
   CodeTD (ours) & $\bf{73.5}$ & $56.9$ \\
   \hline
\end{tabular}
\caption{Transferability of code hallucination detectors, ROC-AUC. Each classifier was trained on HE (MBPP) dataset and tested on MBPP (HE) dataset.}
\label{tab:transfer_he_vs_mbpp}
\vskip -15pt
\end{table}

\subsection{Ranking Ability}
\label{sec:ranking}
Next, we assess the usefulness of CodeTD for ranking code generations. For each problem, all generations were ranked according to the predicted probability of correctness and one with the highest probability was selected. Following pass@1 computation, a baseline was random selection of a code generation.
The use of CodeTD in this setting leads to a significantly higher pass@1 score for all Code LLMs, see Table \ref{tab:ranking_pass1}. This result justifies that the proposed CodeTD  method can be successfully applied for best-of-N sampling.

\begin{table}[t]
\begin{adjustbox}{width=\columnwidth}
\centering
\begin{tabular}{lcc}
    \hline
    \textbf{Model} & \textbf{Random} & \textbf{Clf. Prob.} \\ \hline
    \multicolumn{3}{c}{HE} \\ \hline
    StarCoder2-7B & $28.6 \pm 5.5$ & $\mathbf{43.3 \pm 9.0}$ \\
    CodeLlama-7B & $26.0 \pm 5.1$ & $\mathbf{39.7 \pm 7.2}$ \\
    DeepSeek-Coder-6.7B & $39.1 \pm 4.9$ & $\mathbf{56.7 \pm 7.4}$ \\
    Qwen2.5-Coder-7B & $51.8 \pm 8.0$ & $\mathbf{64.0 \pm 7.3}$ \\
    Magicoder-S-DS-6.7B & $72.5 \pm 10.0$ & $\mathbf{74.3 \pm 6.1}$ \\ \hline
    \multicolumn{3}{c}{MBPP} \\ \hline
    StarCoder2-7B & $43.0 \pm 3.6$ & $\mathbf{49.6 \pm 4.6}$ \\
    CodeLlama-7B & $35.2 \pm 3.3$ & $\mathbf{43.6 \pm 3.4}$ \\
    DeepSeek-Coder-6.7B & $53.0 \pm 2.5$ & $\mathbf{61.4 \pm 2.3}$ \\
    Qwen2.5-Coder-7B & $52.6 \pm 3.6$ & $\mathbf{62.0 \pm 2.4}$ \\
    Magicoder-S-DS-6.7B & $61.4 \pm 3.4$ & $\mathbf{64.8 \pm 2.1}$ \\ \hline
\end{tabular}
\end{adjustbox}
\caption{pass@1 scores across variants of ranking of code generations.}
\label{tab:ranking_pass1}
\end{table}

\subsection{Transferability}

\textbf{Cross-benchmark transferability.} 
In this setting, hallucination classifiers for a fixed Code LLM were trained on data for one benchmark (HumanEval, MBPP) and evaluated on another, then repeated vice versa.
Table \ref{tab:transfer_he_vs_mbpp} demonstrates that CodeTD is transferable, although performance is lower than when training and testing are done on the same benchmark.
The proposed attention features achieve better transferability in 80\% of cases, as measured by ROC-AUC for both the HE $\to$ MBPP and MBPP $\to$ HE transfer. We hypothesize that the changes in the classifiers' performance compared to results of Section \ref{sec:clf_quality} are related to differences in the prompt structure: we use 0-shot prompt for HumanEval and 1-shot prompt for MBPP. The proposed CodeTD has consistent quality when the linguistic complexity of the prompt is increased, see Table \ref{tab:complex_prompt} in Appendix.

\textbf{Cross-model transferability.} We analyze the hallucination detection quality when code-generating and feature-extracting models are different. For each code-generating model, we use several Code LLMs to extract attention features for the generated candidate solutions, see Table~\ref{tab:cross_model}. The best performance across the majority of code-generating models is achieved by Magicoder-S-DS-6.7B, which outperformed a larger DeepSeek-Coder-33B model. On the contrary, the worst detection quality is consistently obtained with StarCoder2-7B model. This result demonstrates that medium-sized models can be powerful feature extractors in code hallucination detection.

\begin{figure}[tb]
\centering
\includegraphics[width=\columnwidth]{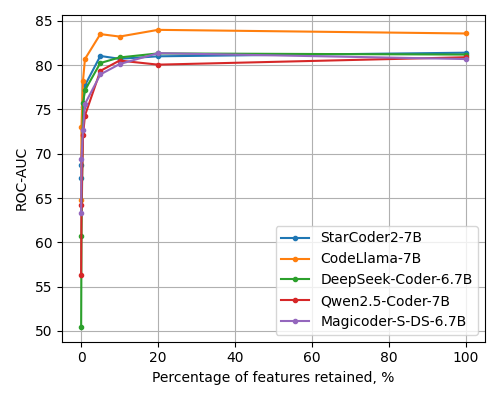}
  \caption{ROC-AUC vs. percentage of retained features, MBPP.}
\label{fig:pruning_rocauc}
\vskip-10pt
\end{figure}

\subsection{Feature Importance}
\label{sec:feat_imp}

In its base setup, CodeTD extracts features from the attention maps of all layers and heads. However, we observed that the trained XGBoost classifier exhibited natural sparsity -- only about 25\% of features were meaningful, based on XGBoost’s feature importance scores.
To further investigate feature importance and selection, we adopted a two-stage pipeline. First, for a given sparsity level, we selected the most critical features according to the importance scores from a classifier trained on all attention features simultaneously. Second, we trained a new XGBoost classifier using only the selected features. As shown in Figure~\ref{fig:pruning_rocauc}, this procedure retained just 5\% of all attention features with no significant loss in classification quality, underscoring that only a small subset of attention heads is relevant for hallucination detection.

\subsection{A Fine-Grained Error Type Classification}
\label{sec:error_types}

We treat specific Python exception types as fine-grained hallucination categories and train a multi-class XGBoost classifier (Table \ref{tab:error_types}).
The results demonstrate that CodeTD is capable of identifying the majority of errors of different types, consistently across the models. Here is a breakdown of detection accuracy of particular types of errors for CodeLlama-7B: AssertionError: 82.0\%, IndexError: 97.8\%, NameError: 75.7\%, RecursionError: 100\%, SyntaxError: 93.3\%, TypeError: 80.6\%, ValueError: 87.5\%, ModuleNotFoundError, ZeroDivisionError, UnboundLocalError, IndentationError, AttributeError, timed out: 80\%. Some error types were grouped because of a very low frequency. Individual features of some heads can classify logic error vs. other errors, see Appendix \ref{app:heads_error_types}.

\subsection{Ablation Study}

We assess the contribution of each proposed CodeTD feature type through an ablation study. We progressively trained an XGBoost classifier, starting with diagonal attention values and then adding 0-dim and 1-dim MTD features, following the order of computational complexity. As shown in Tables \ref{tab:detection_ablation_he_mbpp}, \ref{tab:transfer_ablation}, adding topological features consistently improved hallucination detection in 92\% of cases and led to more robust, transferable classifiers compared to using diagonal features alone.

\begin{table}
\centering
\begin{adjustbox}{width=0.99\columnwidth}
\begin{tabular}{lcc}
    \hline
   \textbf{Model} & \textbf{Accuracy} & \textbf{F1-Score} \\ \hline
   StarCoder2-7B & $0.7 \pm 0.02$ & $0.68 \pm 0.02$ \\
   CodeLlama-7B & $0.66 \pm 0.02$ & $0.62 \pm 0.02$ \\
   DeepSeek-Coder-6.7B & $0.7 \pm 0.03$ & $0.68 \pm 0.04$ \\
   Qwen2.5-Coder-7B & $0.64 \pm 0.02$ & $0.6 \pm 0.02$ \\
   Magicoder-S-DS-6.7B & $0.73 \pm 0.02$ & $0.71 \pm 0.02$ \\ \hline
\end{tabular}
\end{adjustbox}
\caption{Performance of multi-classif. of error types.}
\label{tab:error_types}
\end{table}

\section{Conclusion}

In this paper, we propose CodeTD, a new hallucination detection approach for Code LLMs. 
Our approach provides pre-execution code verification and is useful when functional tests are (a) unavailable, (b) incomplete, (c) expensive, or (d) unsafe to run due to security concerns.
Our approach is based on the introspection of an LLM: we get attention maps, transform them to weighted graphs, and study prompt-generation connectivity patterns.
The proposed topological features of these graphs have been empirically shown to be relevant for code hallucination detection. 
CodeTD outperformed recent baselines in the task of code hallucination detection. CodeTD is transferable across the coding benchmarks and Code LLMs.

\textbf{Application of CodeTD.}
Beyond hallucination detection, the natural next step is hallucination mitigation.
We envision three deployment pathways: (1) best-of-N sampling by picking a sample with the highest CodeTD score (improves pass@1 up to 17.6\%, Section~\ref{sec:ranking}); (2) tracing hallucination up to the first wrong token and regenerating from this step, a natural scenario for an agentic loop; (3) abstention/fallback mechanisms for human-in-the-loop workflows. We leave these directions for future work. 

We believe that our work facilitates the safer deployment of Code LLMs in security-critical applications by providing a pre-execution verification layer.
In a wider context, our work contributes to the study of interpretation and generalization in NLP models, since hallucinations and generalization ability are intrinsically tied.

\section*{Limitations}

Although we have achieved good experimental results, we realize that our research has several limitations.
Our method targets hallucinations that manifest in the geometry of the model’s attention; it does not unravel all root causes (e.g., spurious pre‑training correlations, decoding drift, RLHF bias). Extending the analysis to these factors is left for future work.
Finally, our approach can predict whether a code is correct as a whole but can not point to a specific place with a bug.

\section*{Acknowledgments}
The work was supported by the grant for research centers in the field of AI provided by the
Ministry of Economic Development of the Russian Federation in accordance with the agreement
000000C313925P4F0002 and the agreement №139-10-2025-033.

\bibliography{custom}

\appendix

\section{Details on Generation Procedure}
\label{app:gen_procedure}

\begin{figure}[t]
\begin{center}
  \includegraphics[width=\columnwidth]{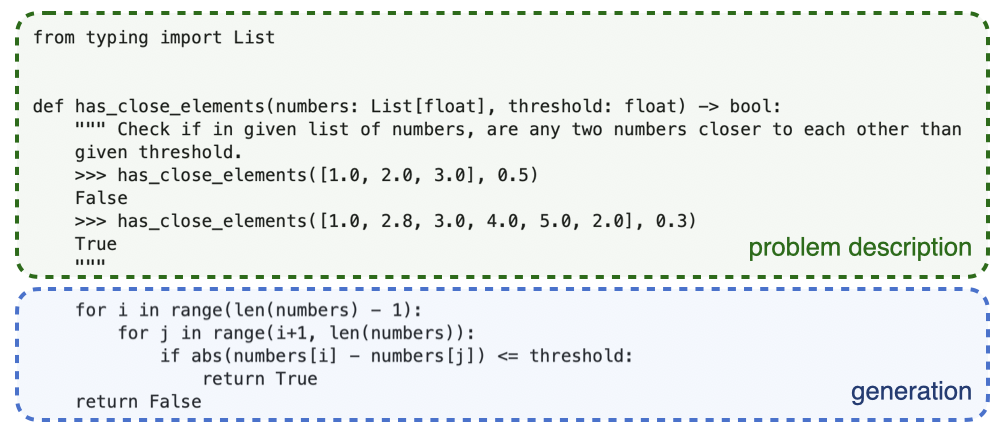}
  \caption{Example of prompt (problem description) and model generation for the HumanEval dataset.}
  \label{fig:he_gen}
\end{center}
\end{figure}

\begin{figure}[t]
\begin{center}
  \includegraphics[width=\columnwidth]{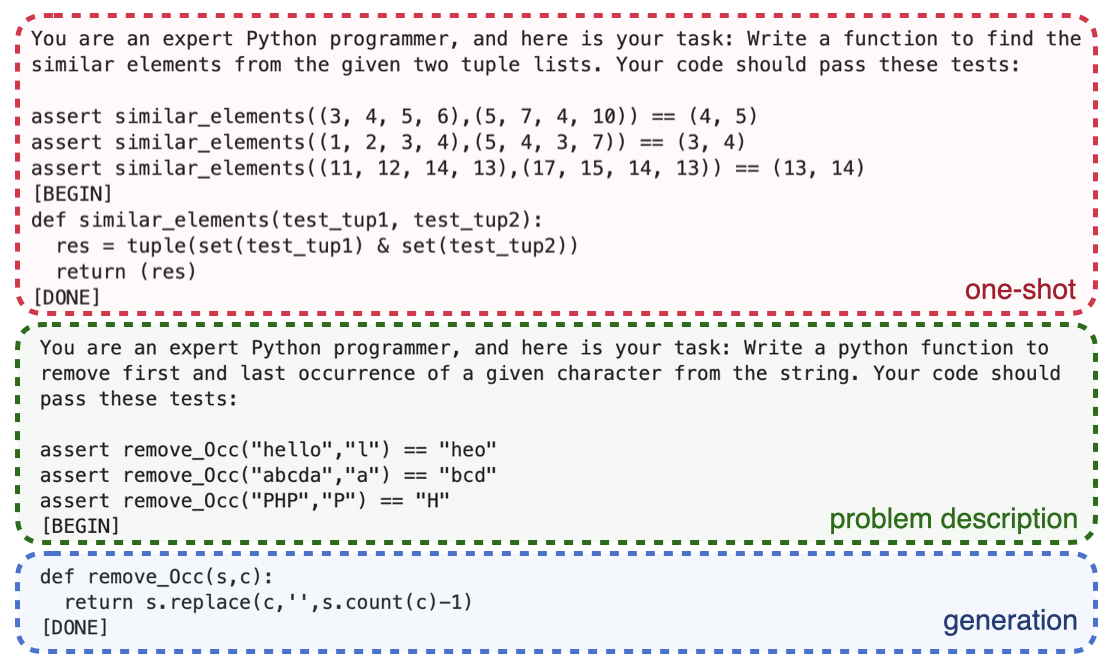}
  \caption{Example of prompt (one-shot example and problem description) and model generation for the MBPP dataset.}
  \label{fig:mbpp_gen}
\end{center}
\end{figure}

We generate solutions for the coding problems with a temperature of $0.8$. For the HumanEval and MultiPL-E datasets, the maximum length of the model output (i.e., input prompt + generation) was limited to $512$ tokens. For the MBPP dataset, the maximum number of new tokens to generate was set to $256$. Figures \ref{fig:he_gen}, \ref{fig:mbpp_gen} provide examples of prompts and generations for HumanEval and MBPP datasets. We followed the guidelines\footnote{https://github.com/bigcode-project/bigcode-evaluation-harness} to post process the model output and extract the valid problem solution. For the BCB dataset, we strictly follow the original generation procedure available at the official repository\footnote{https://github.com/bigcode-project/bigcodebench}. To compute the attention features according to the method proposed in Section \ref{sec:methods}, we used the attention submatrix corresponding to the input prompt and the valid solution to the problem.
Experiments with each dataset-Code LLM took $\approx$ 8 GPU hours.
For computational experiments, we used NVIDIA TITAN RTX. 

\begin{figure*}[t]
\begin{center}
\begin{subfigure}[t]{0.49\textwidth}
\includegraphics[width=\columnwidth]{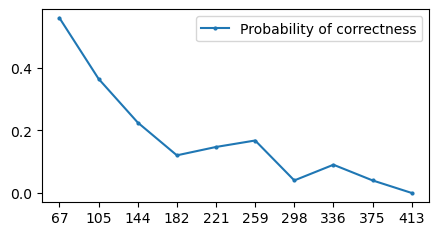}
  \caption{Prompt length, tokens. HumanEval.}
\end{subfigure}
\begin{subfigure}[t]{0.49\textwidth}
\includegraphics[width=\columnwidth]{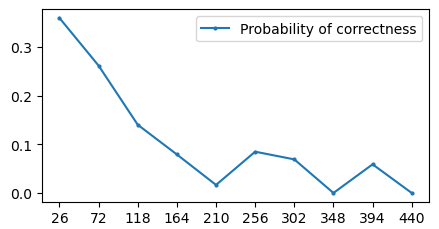}
  \caption{Generation length, tokens. HumanEval.}
\end{subfigure}
\caption{The individual conditional expectations for prompt and generation lengths, CodeLlama-7B.}
\label{fig:promt_gen_vs_correctness}
\end{center}
\end{figure*}

\begin{figure*}[t]
\begin{center}
\begin{subfigure}[t]{0.49\textwidth}
\includegraphics[width=\columnwidth]{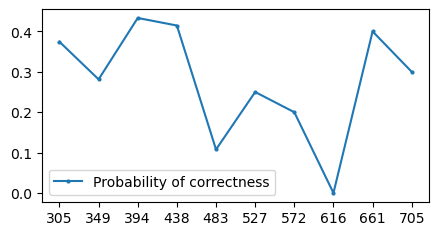}
  \caption{Prompt length, tokens. MBPP.}
\end{subfigure}
\begin{subfigure}[t]{0.49\textwidth}
\includegraphics[width=\columnwidth]{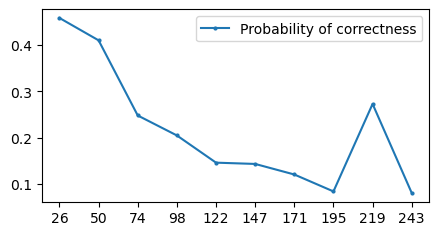}
  \caption{Generation length, tokens. MBPP.}
\end{subfigure}
\caption{The individual conditional expectations for prompt and generation lengths, CodeLlama-7B.}
\label{fig:promt_gen_vs_correctness_mbpp}
\end{center}
\end{figure*}

\section{Details on Training Procedure}
\label{app:train_details}

In CodeTD, we utilized the XGBClassifier with an approximation tree method ``hist'' from the XGBoost library\footnote{https://xgboost.readthedocs.io/en/latest/index.html}.
For the code hallucination detector based on CodeT5-base embeddings, we used the pre-trained frozen CodeT5-base encoder with a trainable classification head consisting of $2$ linear layers with hidden dimensionality $768$. The classification head was trained for $100$ epochs with batch size $32$ and learning rate $3e-5$.

Self-Eval \citep{zhang2024self} is a way to evaluate the responses of an LLM using its internal knowledge. Self-Eval extracts a list of atomic claims from the responses and then prompts an LLM itself to validate the factuality of the claims. Self-Eval is not directly applicable to Code LLMs, since there are no “facts” in the code. However, we applied the core idea of Self-Eval by prompting Code LLMs to evaluate the functional correctness of a generated code. 

In addition, we have adapted Interrogate-LLM \citep{yehuda2024interrogatellm} to detect hallucinations in LLM-generated code. As an embedding model, we used the CodeT5+ 110M Embedding model, $K=5$ and a fixed temperature.

For comparison with attention-based hallucination detectors (AttnLogDet, AttnEigvals, LapEigvals), we followed the procedure in \citet{binkowski2025hallucination} and utilized the code\footnote{https://github.com/graphml-lab-pwr/lapeigvals} to compute attention features. For fair comparison, we keep the number of top eigenvalues to be equal to the number of proposed CodeTD features and use XGBoost as a classifier.

\section{Evaluation Metrics}

This section briefly introduces the evaluation metrics utilized throughout the work.

As the main evaluation metric, we use ROC-AUC to account for possible class imbalance. The \textit{ROC curve} demonstrates the quality of a binary classifier for all possible classification thresholds. The X-axis corresponds to the \textit{False Positive Rate (FPR)} and the Y-axis corresponds to the \textit{True Positive Rate (TPR)} which can be defined as follows: $FPR = \frac{FP}{FP+TN}$,
$TPR = \frac{TP}{TP+FN}$, where \textit{TP} – true positive samples, \textit{FP} – false positive samples, \textit{TN} – true negative samples, \textit{FN} – false negative samples. \textit{ROC-AUC} is defined as the area under the ROC curve. ROC-AUC of a random model is equal to 0.5, ROC-AUC of a perfect model is 1. 

\textit{Accuracy} measures the proportion of correctly classified objects out of the total number of examples:
$$
Accuracy = \frac{TP + TN}{TP + TN + FP + FN}.
$$

The \textit{F1-score} is a harmonic mean of \textit{Precision} and \textit{Recall}:
$$
F_1 = \frac{2}{\frac{1}{Precision} + \frac{1}{Recall}},
$$
where
$$
Precision = \frac{TP}{TP + FP}, Recall = \frac{TP}{TP + FN}.
$$
In multiclass classification, we used weighted average across classes.

To study the ranking ability of the hallucinations detector, we used the \textit{pass@1} metric. pass@1 is a proportion of coding problems from a benchmark for which a Code LLM generated the correct solution passing all the tests, with the restriction that only one solution
is executed.

\section{Examples of Cross-Barcodes}
\label{app:barcodes}

For the MTD feature which strongly distinguishes distribution of classes (Figure \ref{fig:attn_features_codellama}, (a), left), we show Cross-Barcodes having high and small values of this feature, see Figure \ref{fig:mtd_1388}.
See corresponding code samples in Appendix \ref{app:code_samples}.
Note, that the number of bars (the same as number of tokens in generation) is not a distinguishing statistic; the MTD feature is the average length of a bar. Thus, our MTD features do not have spurious correlations with the length of generated code.

Figure \ref{fig:crossbarcodes_4_18} shows examples of Cross-Barcode$_0$ for a fixed attention head. The Cross-Barcode$_1$(P, G) is empty for these attention maps. Correct generations (a), (b) tend to have more and more H$_0$ bars than not-correct ones (c), (d).

Figure \ref{fig:crossbarcodes_15_5} shows examples of Cross-Barcode$_0$, Cross-Barcode$_1$ for a fixed attention head. Correct generations (a), (b) tend to have more and longer H$_1$ bars than not-correct ones (c), (d). 

Attention maps for the corresponding heads are shown in Figures \ref{fig:attmap_4_18}, \ref{fig:attmap_15_5}.
%
%
\begin{figure*}[t]
\begin{center}
\begin{subfigure}[t]{0.45\textwidth}
\includegraphics[width=\textwidth, height=50mm]{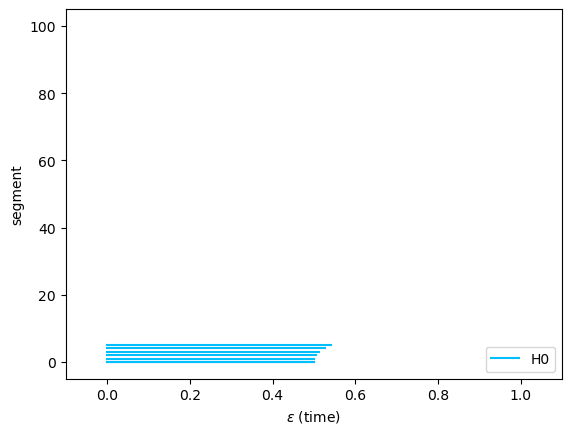}
\caption{Generation 1 (correct).}
\end{subfigure}
\begin{subfigure}[t]{0.45\textwidth}
\includegraphics[width=\textwidth, height=50mm]{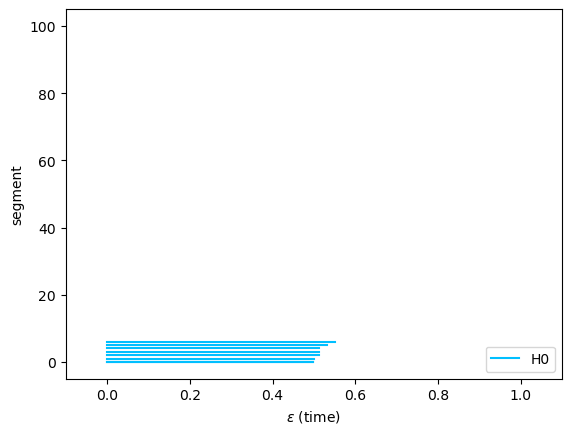}
\caption{Generation 2 (correct).}
\end{subfigure}
\begin{subfigure}[t]{0.45\textwidth}
\includegraphics[width=\textwidth, height=50mm]{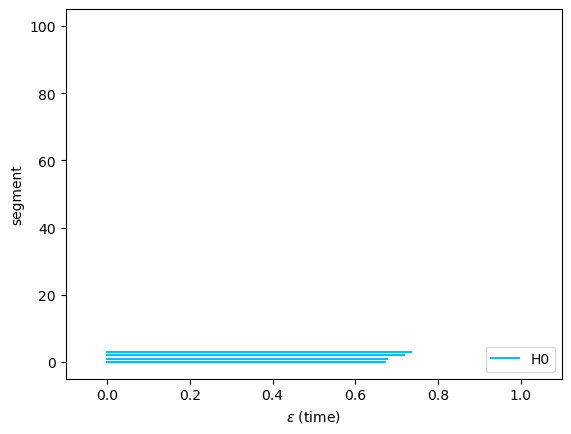}
\caption{Generation 3 (not correct, hallucination).}
\end{subfigure}
\begin{subfigure}[t]{0.45\textwidth}
\includegraphics[width=\textwidth, height=50mm]{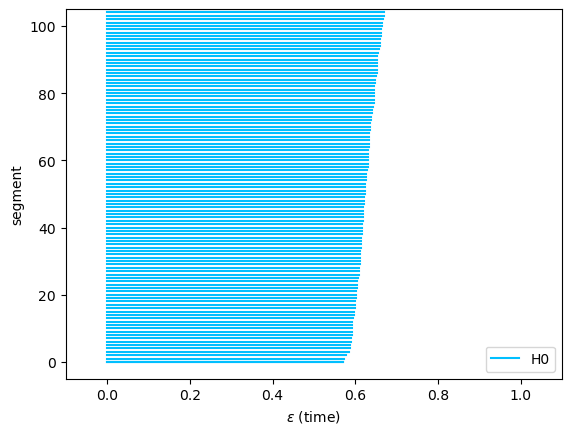}
\caption{Generation 4 (not correct, hallucination).}
\end{subfigure}
\caption{Examples of Cross-Barcode$_0$(P, G), CodeLlama-7B, HumanEval dataset, layer 1, head 17. The number of bars (the same as number of tokens in generation) is not a distinguishing statistic (see Generation 4). The distinguishing statistic is the average length of a bar (normalized MTD). }
\label{fig:mtd_1388}
\end{center}
\end{figure*}
%
%
\begin{figure*}[t]
\begin{center}
\begin{subfigure}[t]{0.45\textwidth}
\includegraphics[width=\textwidth, height=50mm]{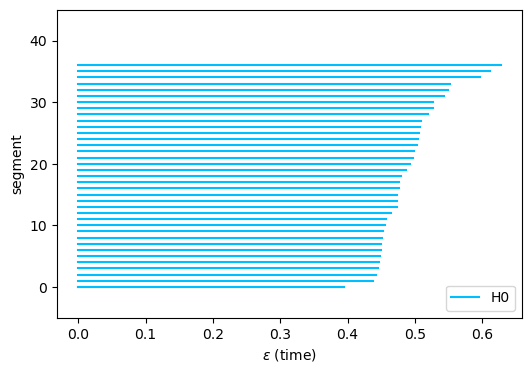}
\caption{Generation 1 (correct).}
\end{subfigure}
\begin{subfigure}[t]{0.45\textwidth}
\includegraphics[width=\textwidth, height=50mm]{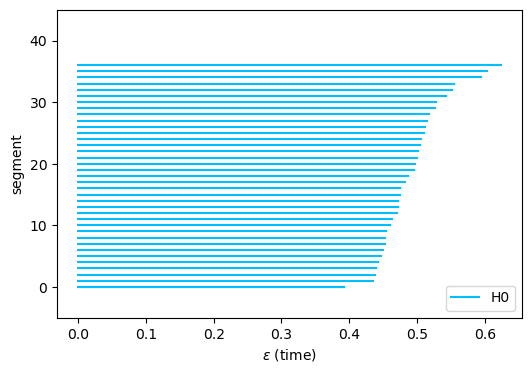}
\caption{Generation 2 (correct).}
\end{subfigure}
\begin{subfigure}[t]{0.45\textwidth}
\includegraphics[width=\textwidth, height=50mm]{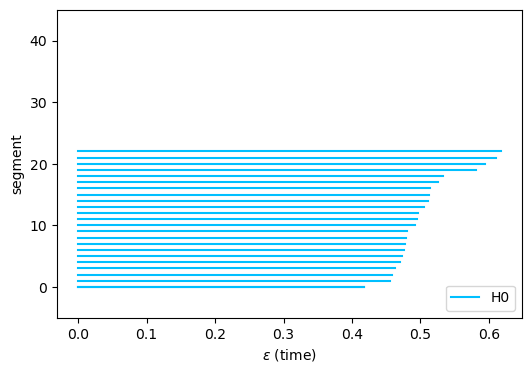}
\caption{Generation 3 (not correct, hallucination).}
\end{subfigure}
\begin{subfigure}[t]{0.45\textwidth}
\includegraphics[width=\textwidth, height=50mm]{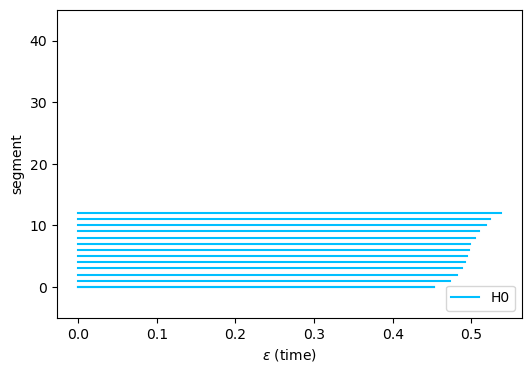}
\caption{Generation 4 (not correct, hallucination).}
\end{subfigure}
\caption{Examples of Cross-Barcode$_0$(P, G), CodeLlama-7B, HumanEval dataset, problem 14, layer 4, head 18. Cross-Barcode$_1$(P, G) is empty for these attention maps.}
\label{fig:crossbarcodes_4_18}
\end{center}
\end{figure*}
%
%
\begin{figure*}[t]
\begin{center}
\begin{subfigure}[t]{0.45\textwidth}
\includegraphics[width=\textwidth, height=45mm]{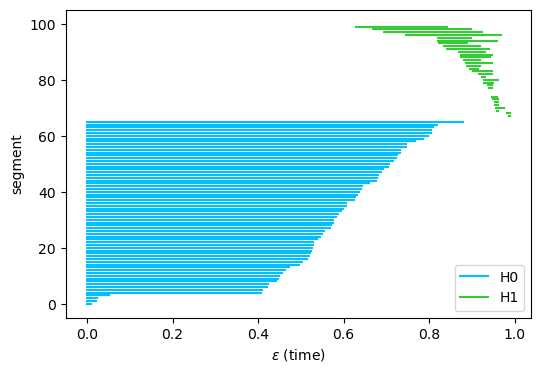}
\caption{Generation 1 (correct).}
\end{subfigure}
\begin{subfigure}[t]{0.45\textwidth}
\includegraphics[width=\textwidth, height=45mm]{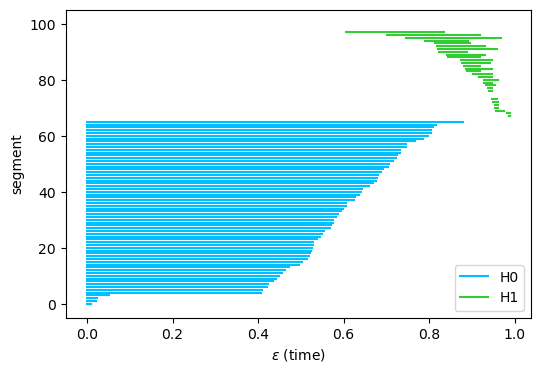}
\caption{Generation 2 (correct).}
\end{subfigure}
\begin{subfigure}[t]{0.45\textwidth}
\includegraphics[width=\textwidth, height=45mm]{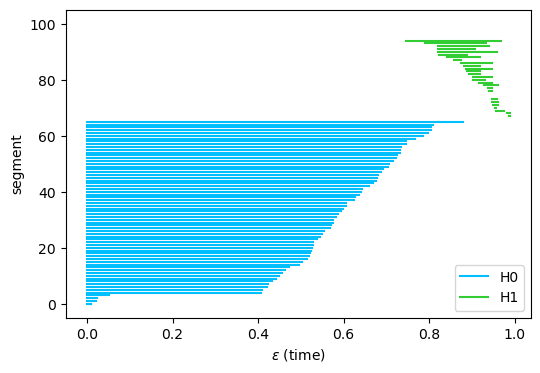}
\caption{Generation 3 (not correct, hallucination).}
\end{subfigure}
\begin{subfigure}[t]{0.45\textwidth}
\includegraphics[width=\textwidth, height=45mm]{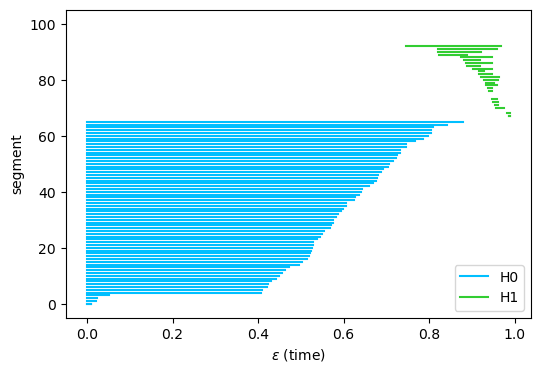}
\caption{Generation 4 (not correct, hallucination).}
\end{subfigure}
\caption{Examples of Cross-Barcode$_0$(G, P), Cross-Barcode$_1$(G, P). CodeLlama-7B, HumanEval dataset, problem 14, layer 15, head 5.}
\label{fig:crossbarcodes_15_5}
\end{center}
\end{figure*}
%
%
\begin{figure*}[t]
\begin{center}
\begin{subfigure}[t]{0.45\textwidth}
\includegraphics[width=\textwidth, height=55mm]{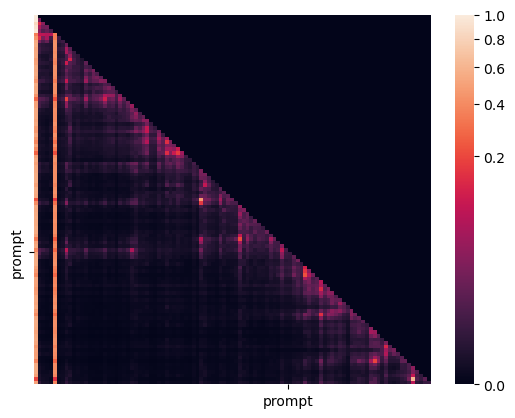}
\caption{Generation 1 (correct).}
\end{subfigure}
\begin{subfigure}[t]{0.45\textwidth}
\includegraphics[width=\textwidth,  height=55mm]{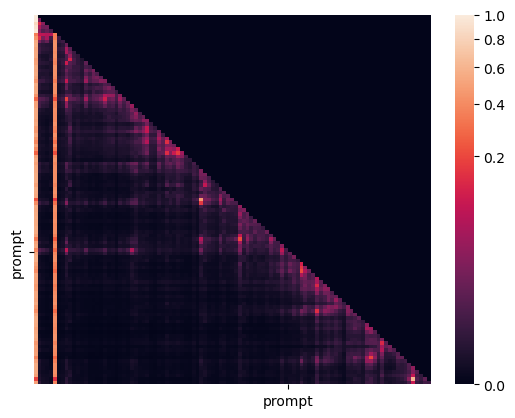}
\caption{Generation 2 (correct).}
\end{subfigure}
\begin{subfigure}[t]{0.45\textwidth}
\includegraphics[width=\textwidth,  height=55mm]{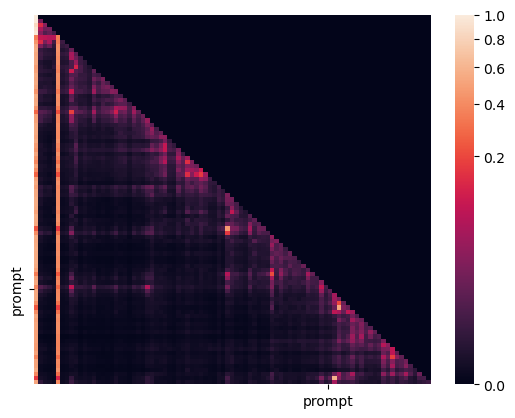}
\caption{Generation 3 (not correct, hallucination).}
\end{subfigure}
\begin{subfigure}[t]{0.45\textwidth}
\includegraphics[width=\textwidth,  height=55mm]{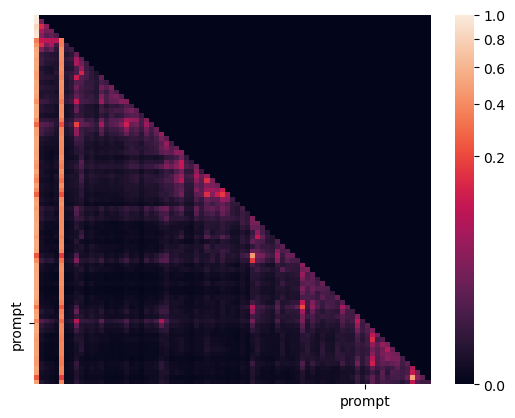}
\caption{Generation 4 (not correct, hallucination).}
\end{subfigure}
\caption{Attention maps. CodeLlama-7B, HumanEval dataset, problem 14, layer 4, head 18.}
\label{fig:attmap_4_18}
\end{center}
\end{figure*}
%
%
\begin{figure*}[t]
\begin{center}
\begin{subfigure}[t]{0.45\textwidth}
\includegraphics[width=\textwidth, height=45mm]{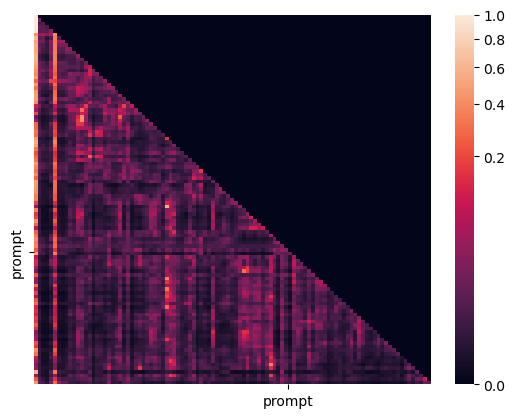}
\caption{Generation 1 (correct).}
\end{subfigure}
\begin{subfigure}[t]{0.45\textwidth}
\includegraphics[width=\textwidth, height=45mm]{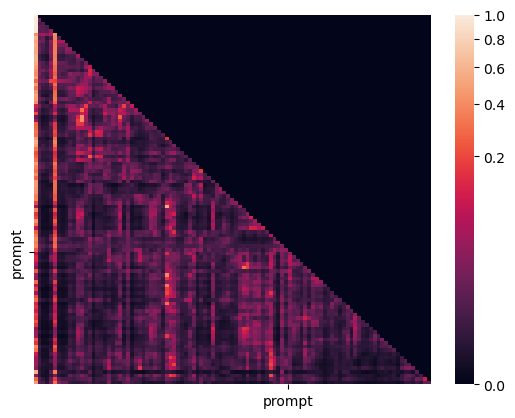}
\caption{Generation 2 (correct).}
\end{subfigure}
\begin{subfigure}[t]{0.45\textwidth}
\includegraphics[width=\textwidth, height=45mm]{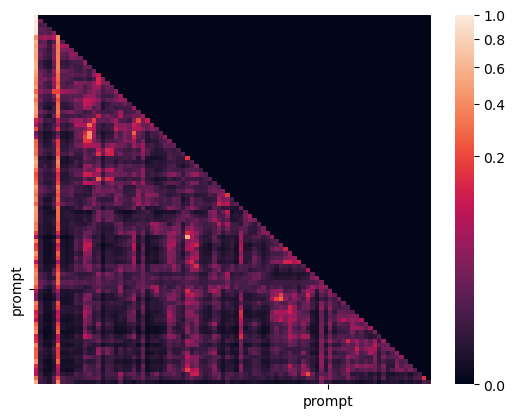}
\caption{Generation 3 (not correct, hallucination).}
\end{subfigure}
\begin{subfigure}[t]{0.45\textwidth}
\includegraphics[width=\textwidth, height=45mm]{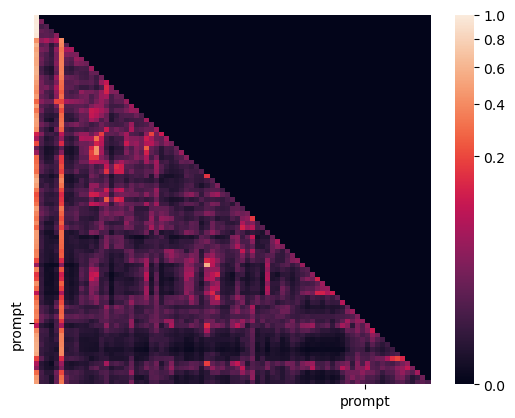}
\caption{Generation 4 (not correct, hallucination).}
\end{subfigure}
\caption{Attention maps. CodeLlama-7B, HumanEval dataset, problem 14, layer 15, head 5.}
\label{fig:attmap_15_5}
\end{center}
\end{figure*}

\section{Ablation for the Classifier Model Selection}\label{app:clf_ablation}

We carried out additional experiments with the feed-forward network (MLP), logistic regression, and support vector classifier (SVC) instead of XGBoost as a classifier for hallucination detection (the rightmost block in Figure \ref{fig:pipeline}). We used MLP with two hidden layers of size 256 and ReLU activations. This configuration was selected after moderate optimization of an architecture. For logistic regression and SVC, we tuned the value of regularization strength. Table \ref{tab:he_mbpp_mlp_vs_xgboost} presents the results.
XGBoost offers (a) strong average performance across all code LLMs, (b) negligible training cost ($\approx$ 30 s per fold), and (c) no hyper‑parameter tuning in our setting. XGBoost guarantees low computational overhead while providing a single, robust baseline for subsequent work. This experiment demonstrates that the high performance of the proposed method is caused by the relevance of the extracted attention features and not by a specific choice of classifier.

\section{Applicability of GNNs to Attention Maps}

To train a GNN-based approach on graphs with edge weights obtained from attention matrices, these attention matrices need to be stored. We can estimate the approximate memory footprint to store attention matrices of size (seq\_len\_k)$^2$ for a model with n\_layers and n\_heads for a dataset of size N using the formula: n\_layers $\times$ n\_heads $\times$ s $\times$ $\sum_{i=1}^k$
 (seq\_len\_k)$^2$ where s is the size of the float type. We assume s = 4 bytes. If one uses only attention matrices from the last layer of the model, we obtain the memory footprint approximately 20.7 - 31.1 GB for the Human Eval dataset and 36.6 - 53.7 GB for MBPP (depending on the model). However, to store the attention matrices for all layers and all heads, the memory footprint is about 578.2 - 996.4 GB for Human Eval and 1023.5 - 1718.7 GB for MBPP. We highlight that even for datasets of moderate size (i.e. 4100 generations for HumanEval and 2500 generations for MBPP), the memory footprint becomes prohibitively high. Thus, it is not always feasible to store such features. In contrast, in our approach, we do not need to store the attention matrices since we compute all the features immediately during generation. Hence, the size of our training dataset is negligible. Moreover, our approach demonstrates high performance without hyperparameter tuning. Therefore, we believe that the proposed approach has better scalability and is more practical.

\section{Computational Complexity}
\label{app:complexity}
 
We provide an estimate of computational time using the CodeLlama-7B model and HumanEval dataset as an example. The average time taken to generate a solution for one problem without CodeTD is 6.2 sec, with CodeTD 11.5 sec, when topological features are calculated for all layers and heads, which are processed in parallel. Feature computation time during inference can be further decreased if only the most important features are used for classification: according to Section \ref{sec:feat_imp}, it is possible to use only 5\% of all attention features without significant loss of classification quality. The average memory footprint is $\approx$190MB for HumanEval and $\approx$544MB for MBPP. This size is a small fraction of GPU footprint during generation.
 
\section{On Definition of a Code Hallucination}
\label{app:hallucination_definition}

In our approach, the topological features obtained from attention maps account for the dissimilar structures in the prompt and generation subsets. Our intuition is that a correct solution should correspond to the structure of the prompt as their high-level semantic meanings cohere. Although other reasons behind hallucinations are possible, our approach estimates the correctness of code based only on the internal information flow of the model that does not require additional resources. Nevertheless, the proposed approach can be further integrated into other code hallucination detection tools to achieve better performance. Also, the most popular benchmarks like HumanEval and MBPP check only functional correctness by running functional tests, that is, whether a code solves the corresponding problem as stated in a prompt.

\textbf{Hallucinations and generalization}. The study of hallucinations in LLMs is intrinsically tied to generalization in NLP models. Both challenges stem from the way models learn, represent, and apply knowledge. Improving generalization through robust training, diverse data, and better uncertainty handling reduces hallucinations by ensuring that the models produce contextually appropriate and factually grounded output. In contrast, analyzing hallucinations provides an insight into generalization failures, guiding the development of more reliable NLP systems. This symbiotic relationship underscores the importance of addressing both issues holistically in AI research.

\textbf{MTD as manifold-topology divergence in value space (per head)}. For a decoder-only attention head, the output representation at token (i) is an attention-weighted mixture of value vectors $o_i=\sum_{j\le i} a_{ij}v_j$, 
, so the attention coefficients determine how tokens are embedded into the attended value-representation space of that head. Moreover, in decoder models the attention coefficients can be viewed (under the standard approximation used in our intuition) as tracking similarity structure induced by scalar products among the corresponding token representations that drive value mixing; under norm-concentration / approximate isotropy assumptions (vectors lying on a sphere of roughly fixed radius), this motivates treating $w_{ij}=1-a_{ij}$
 as an angular-dissimilarity-like transform for the filtration. In this sense, the Cross-Barcode / $\mathrm{MTD}_0(P,G)$
 construction can be interpreted as measuring a multiscale topological divergence between the latent manifolds underlying the prompt-token point cloud (P) and the generation-token point cloud (G) in the head’s attended value space (i.e., how much scale is needed before generation structures become topologically connected to prompt structures).

\textbf{MTD as an area-under-the-curve multiscale integration cost}. An equivalent and intuitive $H_0$ interpretation is to define $U_{\mathrm{comp}}(\tau)$
 as the number of connected components in the filtered attention graph at threshold 
 that contain generation tokens but are not yet connected to the prompt set (P). Then $\mathrm{MTD}_0(P,G)$
 is the area under this curve across thresholds, i.e., the total lifetime of prompt-disconnected generation components over the filtration. This quantity is not only a measure of delayed prompt grounding, but also a multiscale integration cost: it accumulates how long generation-side components remain disconnected from the prompt-conditioned structure and from one another before merging as the threshold grows. Hence, small $\mathrm{MTD}_0$
 indicates early integration (globally coherent, prompt-aligned generation), whereas large $\mathrm{MTD}_0$
 indicates persistent fragmentation and delayed integration, which is empirically associated with functional incorrectness.

\textbf{Stability}. By \cite{cohen2005stability}, Cross-Barcodes satisfy $d_B(\mathbf{B}(W), \mathbf{B}(W')) \le |W - W'|_\infty$. Hence $MTD_k$ inherits Lipschitz stability under attention perturbations --- a property \textit{not enjoyed} by spectral baselines, since eigenvalue ordering can be unstable under near-degeneracies. This is, in our view, a key reason CodeTD outperforms LapEigenvals on transferability (Table 3, 80\% of cells).

\section{Experiments with Larger Models}
\label{app:large_models}

First, Table \ref{tab:detection_results_he_large} demonstrates CodeTD outperforms other baselines even when the model ability to generate correct code candidates increases.

\textbf{DeepSeek R1 671B.} We carried out additional experiments, where code hallucinations are detected by a recent reasoning DeepSeek R1 model having $671$ billion parameters with Chain of Thought inference on MBPP dataset. We used the following prompt:

\textit{You are provided with two coding tasks with solutions. The first one is just an example and does not need an assessment.
Tell whether the second task is correctly solved by the code provided in the second [BEGIN] [DONE] block. The answer must be Yes or No}

For code generated with the DeepSeek-6.7B model, we asked DeepSeek R1 using the above prompt, extracted the final answers (i.e. ``Yes'' or ``No'') from the generated responses, and trained the XGBoost classifier using these features. The quality of hallucination detection via such zero-shot prompting is in Table \ref{tab:detection_deepseek_r1}, row ``DeepSeek R1, (671B model)''. The quality of the proposed approach is shown in Table \ref{tab:detection_deepseek_r1}, row ``CodeTD (ours, from 6.7B model)''. Note that in this case, we use \textit{only} the attention features of the DeepSeek-6.7B model obtained during code generation. Finally, we combine both types of features to train a classifier and report its performance in Table \ref{tab:detection_deepseek_r1}, row ``CodeTD (ours, from 6.7B model) + DeepSeek R1''. The larger DeepSeek R1 model demonstrates better performance than the classifier trained on attention features. However, by adding an output of DeepSeek R1 to our attention-based features and training the XGBoost classifier, we can achieve the best ROC-AUC score. The study of DeepSeek R1's Chain of Thoughts shows that this LLM is doing verification of code by interpreting Python code step by step for unit tests. This can explain the high accuracy of DeepSeek R1.
At the same time, our method opens opportunities for a deeper understanding of inner working and information flow inside transformer models. Our attention features were based on a small 6.7B model in this experiment, however, our features were able to improve the performance of DeepSeek~R1.

\textbf{Qwen2.5-Coder-32B.} Additionally, we performed a similar experiment with Qwen2.5-Coder-32B. In this case, we use the same Qwen2.5-Coder-32B to generate code, extract attention features, and evaluate its performance with zero-shot prompting. Table \ref{tab:detection_qwencoder32b} provides the experimental results. The proposed CodeTD is capable of achieving a better detection quality than the zero-shot prompting, which supports the applicability of the proposed approach to larger models.

\textbf{CodeJudge.} Furthermore, we provide a comparison with CodeJudge \cite{tong2024codejudge}, a code evaluation framework that uses LLM to analyze code functionality and then decide on code correctness. In our comparison, we use CodeLlama-Instruct 34B as an evaluation model to produce binary output to show whether the generated code is correct or not. We report the mean results over 3 runs in accordance with the original setup; see Table \ref{tab:codejudge}.

We highlight that for a fair comparison we should consider the setup when CodeJudge is run without reference and CodeTD is run for large 32-34B models. In our work, we did not include reference (i.e. correct solution) to the prompt as this setup is not practical (typically, one does not know the correct solution). Also, since CodeJudge is evaluated with a 34B model, we evaluate CodeTD in a similar way to keep the experimental design of both methods as close as possible. In this comparison, CodeTD outperforms CodeJudge, winning by a large margin for ROC-AUC. However, to further explore the capabilities of the proposed CodeTD, we provide additional comparison when CodeJudge is run with reference and when CodeTD is run for smaller 6.7-7B models. In these cases, CodeTD still outperforms CodeJudge as measured by ROC-AUC.

\section{Contribution of Individual Heads}
\label{app:more_languages}

We carry out additional experiments with 7B-sized models and MultiPL-E benchmark\footnote{https://github.com/nuprl/MultiPL-E}, which is a translation of HumanEval to several popular programming languages; we used Go, Java, Rust, Lua among them.
We find that features of some heads have quite a high correlation with the target value (presence of a hallucination) and can be used as individual predictors. In Table \ref{tab:more_languages} we report the ROC-AUC scores of the top-performing features.

\section{Evaluation with Greedy Decoding}
\label{sec:greedy}

In the main experiments in Section \ref{sec:clf_quality}, we use sampling with temperature $0.8$ to generate diverse solutions for each coding problem and obtain a larger train and test samples. However, production systems often use greedy decoding, and in this section we fill this gap. We evaluate the performance of the proposed CodeTD method when both train and test data are generated with greedy decoding. In this setup, the sample sizes are 164 for HumanEval and 500 for MBPP with 1 generation per task, and we use cross-validation with 5 folds. Although the sample size is sufficiently smaller than in Section \ref{sec:clf_quality}, the proposed CodeTD achieves high performance across all models and datasets and can be applied to small samples; see Table \ref{tab:detection_results_greedy}. The proposed CodeTD consistently outperforms CodeJudge (see Table \mbox{\ref{tab:codejudge}}) both with greedy decoding and sampling. Moreover, while CodeJudge performance may deteriorate when increasing the sample size with sampling, the proposed approach demonstrates stable improvement. 

\section{Evaluation with Partial Generation}
\label{sec:part_gen}

Evaluating very long token sequences in a single pass can be computationally prohibitive, yet processing the full sequence is necessary to obtain the complete feature set. To address this limitation, we evaluated our method using a sliding-window generation strategy. Specifically, the input sequence is divided into N chunks, each of size at most L. Each chunk is passed independently through the model, after which attention maps are collected and features are computed. The per-chunk features are then concatenated into a single feature vector and passed to XGBoost. The results for DeepSeek-Coder-6.7B are reported in Table \ref{tab:part_gen}. These results demonstrate that the sliding-window approach substantially reduces memory consumption, with some loss in accuracy.

\begin{table}[]
    \centering
    \begin{tabular}{ll}
        \hline
        \textbf{Part} & \textbf{ROC-AUC} \\ \hline
        100 tokens & 0.633 ± 0.034 \\
        300 tokens & 0.673 ± 0.012 \\
        500 tokens & 0.656 ± 0.048 \\
        Full generation & 0.69 ± 0.02 \\ \hline
    \end{tabular}
    \caption{CodeTD performance on partial generation}
    \label{tab:part_gen}
\end{table}

\section{Performance on Different Task Lengths}

We evaluated our method on different problem lengths. For this purpose we stratified problems by prompt length (small: <300, medium: 300–500, long: >500 tokens) and evaluated CodeTD (default settings, stratified 5‑fold CV). CodeTD achieved higher mean ROC-AUC across the majority of Code LLM length categories. The evidence supports the conclusion that CodeTD maintains consistent performance across all prompt lengths. The results are shown in Table \ref{tab:dif_length_perf}.

CodeTD matches or outperforms LapEigvals across all models except StarCoder2. This isolated result should be interpreted cautiously, as the StarCoder2 dataset is highly imbalanced, with hallucinations comprising only 5.9\% of the observations. Stratification further amplifies this imbalance, and the StarCoder2 long-prompt subset contains only two positive examples, making the corresponding performance estimate statistically fragile.

\begin{table*}
\centering
\begin{tabular}{llll}
    \hline
    \textbf{Method} & \textbf{<300} & \textbf{300–500} & \textbf{>500} \\ \hline
    \multicolumn{4}{c}{DeepSeek-Coder-6.7B} \\ \hline
    LaplEigvals  & 0.600 ± 0.037 & 0.646 ± 0.051 & 0.559 ± 0.109 \\
    CodeTD & \textbf{0.716 ± 0.051} & \textbf{0.697 ± 0.069} & \textbf{0.561 ± 0.160} \\ \hline
    \multicolumn{4}{c}{CodeLlama-7B} \\ \hline
    LaplEigvals & 0.700 ± 0.019 & 0.749 ± 0.067 & 0.738 ± 0.166 \\
    CodeTD & \textbf{0.737 ± 0.033} & \textbf{0.766 ± 0.033} & \textbf{0.755 ± 0.126} \\ \hline
    \multicolumn{4}{c}{Qwen2.5-Coder-7B} \\ \hline
    LaplEigvals & 0.648 ± 0.040 & 0.648 ± 0.089 & \textbf{0.637 ± 0.154} \\
    CodeTD & \textbf{0.697 ± 0.042} & \textbf{0.654 ± 0.087} & 0.584 ± 0.165 \\ \hline
    \multicolumn{4}{c}{Magicoder-S-DS-6.7B} \\ \hline
    LaplEigvals & 0.626 ± 0.053 & \textbf{0.664 ± 0.040} & 0.641 ± 0.101 \\
    CodeTD & \textbf{0.662 ± 0.054} & 0.642 ± 0.037 & \textbf{0.694 ± 0.052} \\ \hline
    \multicolumn{4}{c}{StarCoder2-7B} \\ \hline
    LaplEigvals & \textbf{0.879 ± 0.006} & \textbf{0.833 ± 0.044} & \textbf{0.987 ± 0.013} \\
    CodeTD & 0.831 ± 0.001 & 0.810 ± 0.027 & 0.962 ± 0.013 \\ \hline
\end{tabular}
\caption{CodeTD performance on different sequence lengths.}
\label{tab:dif_length_perf}
\end{table*}

\section{Failure Case Study}

Table \ref{tab:failure_case} provides a failure case study. For each trained classifier, we analyzed the execution results of the generated code snippets which were incorrectly classified by the classifier. We report the fraction of top-3 most popular code categories w.r.t. number of samples in the test sample. We used the 5-fold stratified group cross-validation, and we report the average values over the 5 folds. The table reveals the most popular failure cases: misclassification of correct codes that passed the tests (referred to as ``passed'') and misclassification of wrong solutions (i.e., code snippets that did not pass any test and caused an AssertionError). The fraction of other errors (top-3 and others) is sufficiently lower.

\section{Breakdown by Error Type}
\label{app:error_types}

We consider a Python exception type after running the functional test as a hallucination type. Here are the common exceptions from the HumanEval and MBPP benchmarks: AssertionError, AttributeError, IndentationError, IndexError, ModuleNotFoundError, NameError, RecursionError, SyntaxError, Type Error, UnboundLocalError, ValueError, ZeroDivisionError, timed out.
Most of the errors are due to failing functional tests (AssertionError fraction), see Table \ref{tab:assertion_error_types}.

\section{Individual Features Detect Error Types}
\label{app:heads_error_types}
We carried out additional experiments and found that some of the proposed features can distinguish logical errors (AssertionError) vs. syntax and runtime errors (SyntaxError, ZeroDivisionError, NameError, etc.). Here are the results for CodeLlama-7B, HumanEval dataset:
\begin{itemize}
\item answer diag.attention, layer 29, head 24: ROC\_AUC = 0.67
\item MTD$_0$(P, G) layer 0, head 4: ROC\_AUC = 0.66
\item MTD$_0$(P, G) layer 1, head 10: ROC\_AUC = 0.66
\end{itemize}


\section{Manifold Topology Divergence}
\label{app:mtd}

\begin{figure*}[t]
\centering
  \includegraphics[width=2\columnwidth]{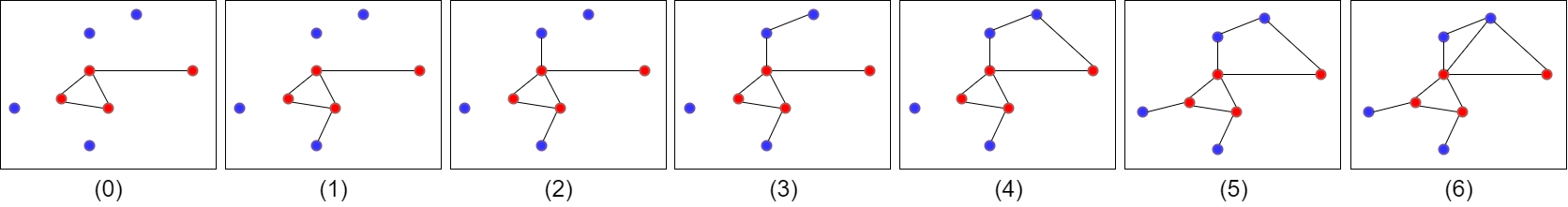}
  \caption{An example of MTD evaluation for a graph having two groups of vertices -- red and blue. (0): initially, only edges connecting red vertices are present. (1)-(6): the rest of edges are added sequentially in an ascending order by their weights. While adding edges, connected components merge with each other. These moments are depicted by $H_0$ bars in Fig. \ref{fig:mtd_barcode}. At moment (4) a cycle appears, at moment (6) this cycle disappears. These moments are depicted by the $H_1$ bar in Fig. \ref{fig:mtd_barcode}.}
  \label{fig:mtd_evolution}
\end{figure*}

MTD (Manifold Topology Divergence) \citep{Barannikov2021manifold} is a tool of TDA that can be used to evaluate the ``dissimilarity'' between two sets of vertices in a weighted graph $\mathcal{G} = (V, E, W)$ or, in other words, to which degree one set of vertices is covered by another set. 

\begin{figure}[tb]
\centering
\includegraphics[width=0.45\textwidth]{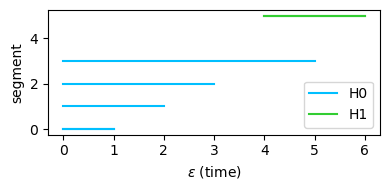}
\caption{Cross-Barcode for a filtration from Fig. \ref{fig:mtd_evolution}.}
\label{fig:mtd_barcode}
\end{figure}

Let a set of vertices $V = P \sqcup G$, be split into disjoint sets $P,  G$.
We consider a nested sequence of graphs $\mathcal{G}_0 \subset \ldots \subset \mathcal{G}_{i} \subset \mathcal{G}_{i+1} \subset \ldots \subset \mathcal{G}$ in the following way.
$\mathcal{G}_0$ has all the vertices $P, G$ and all the edges that connect the vertices of $P$. The sequence $\mathcal{G}_i$ is obtained by adding the rest of the edges one by one in ascending order of their weights; see Figure \ref{fig:mtd_evolution}.
During this process, graph topology naturally changes: connected components are merged, cycles appear and disappear, etc. This process is rigorously described by the theory of \textit{persistence barcodes} \citep{Chazal2019introduction}. Each topological feature, such as connected component or cycle, has a ``birth time'' and a ``death time'', by a corresponding edge weight. The multi-set of these birth-death pairs (intervals) altogether is called a Cross-Barcode$_k$, see Figure \ref{fig:mtd_barcode}. Here $k$ is an index of a \textit{persistence homology}, each of them reflects a kind of topological feature: 0 - connected components, 1 - cycles, 2 - voids, etc.
MTD$_k$ is an integral characteristic of a Cross-Barcode$_k$ and it is defined as a sum of birth-death intervals' lengths 
$$
\text{MTD}_k = \sum_{(b_i, d_i) \in CrossBarcode_k} d_i - b_i.
$$
The higher MTD$_k$ is, the greater is the ``dissimilarity'' between sets of tokens.
Note that according to the definition, MTD$_k$ is not symmetric. 
MTD$_k$, as a kind of persistence barcode, enjoys stability w.r.t. perturbations of weights \citep{cohen2005stability}.

\section{On Importance of Cycles}
\label{app:cycles}

Graph cycles through the prompt graph $P$ are abundant, since $P$ is fully connected.
Cross-Barcode$_k$(P, G) is computed on the clique complex $K(G_k)$, not on the graph itself. Because $K(G_0)$ contains the full simplex on $P$, and the clique complex of a complete graph is contractible, every $P$-internal 1-cycle is already filled by a 2-chain at $k=0$:
\[
\dim H_1(K(G_0)) = 0.
\]
A non-trivial 1-cycle is therefore a \textbf{detour loop}: $p_1 \to g_1 \to \dots g_k \to p_2$, i.e., two prompt nodes joined through a G-path while a path $p_1 \to p_2$ in P is already present. The cycle is ``born'' when the last-detour edge enters the filtration, and cycle ``dies'' when its nodes become pairwise connected. See Fig. \ref{fig:mtd_evolution} for step-by-step illustration for cycles detection. Cycle persistence is therefore how long precisely generation tokens fail to follow the prompt's local geometry: they connect prompt nodes $p_1, p_2$ via a G-path but remain weakly aligned with the prompt themselves. Correct generations bind tightly, detour loops collapse fast, MTD$_1$ is small; incorrect generations drift, detour loops persist, MTD$_0$ is large. 
MTD$_0$ relates to connected components and measures whether G-components attach to P at all; MTD$_1$ measures how \textit{cleanly} they attach.
This is an explanation of results from \cite{Barannikov2021manifold}.

\section{Comparison with TOHA \cite{bazarova2025hallucination} }
\label{app:bazarova}

\citet{bazarova2025hallucination} introduced TOHA, a Topology-based
Hallucination detector, which leverages a topological divergence metric to quantify the structural properties of graphs induced by attention matrices in Retrieval-Augmented Generation (RAG) settings where models may generate outputs unsupported by the provided context. The TDA-on-attention research line on which our work builds --
\citet{Barannikov2021manifold},
\citet{kushnareva2021artificial}, \citet{cherniavskii2022acceptability}, \citet{tulchinskii2022topological} -- is already cited in the manuscript. We welcome the comparison: it sharpens what is novel in CodeTD. 
The central methodological difference is that \citet{bazarova2025hallucination} operate head-by-head: for each attention head, they compute a per-head score, rank heads by their individual discriminative power on a held-out split, and use a small selected subset as the predictor.
This select-then-aggregate pipeline bakes in a strong prior - that signal lives in a few ``best'' heads — and discards the rest of the model. CodeTD takes the opposite stance: extract the complete set of topological features across every layer and head and let a gradient-boosted classifier learn the joint structure. Per head selection is a degenerate special case (a rank-1 sparse classifier on a single slice). Section 7.5 / Fig. 4 show the optimal sparse subset is not ``a few best heads'' but a structured ~5\% pattern across layers, heads, and homological dimensions that XGBoost discovers automatically.
Table \ref{tab:bazarova} presents a detailed comparison.

\begin{table*}[]
    \centering
    \begin{tabular}{p{4cm}p{3.5cm}p{7cm}}
\hline
\textbf{Axis} & \cite{bazarova2025hallucination} & \textbf{CodeTD (Ours)} \\
\hline
Task & RAG hallucination in QA/summarization & Functional incorrectness of code \\
Ground Truth & Text-level factuality & Execution-based labels \\
 Role of the Prompt & Knowledge context & Formal specification \\
MTD Orientation & (P, G) only & Asymmetric: both (P, G) and (G, P) used \\
Homology Dimensions & $H_0$ & $H_0$ and $H_1$ \\
Aggregation & Per-head selection & Joint GBDT over all features \\
Diagonal Self-Attention & — & Included as per-vertex features complementing edge-level topology \\
Models & General LLMs & 10 Code LLMs  (1.5B – 34B) \\
Languages & Natural language & Prompt in natural language; generation in 5 programming languages: Python, Java, Go, Rust, Lua. \\
Cross-language Stability & — & The same predictive (layer, head) features transfer across all 5 programming languages (see Tab. 11). \\
Fine-grained Error Types& — & Multi-class detection of Python exceptions (see Tab. 5, App. N). \\
Ranking of samples & — & pass@1 gains up to +17.6 pp. \\
\hline

    \end{tabular}
    \caption{Detailed comparison with \cite{bazarova2025hallucination}.}
    \label{tab:bazarova}
\end{table*}

Our work provides a theoretical contribution beyond using MTD merely as a feature. Appendix \ref{app:hallucination_definition} develops an interpretation specific to decoder-only transformer attention, viewing MTD as a topological invariant of distribution supports in the head's attended value space. AttnLogDet, AttnEigvals, and LapEigvals (Binkowski et al., 2025) are the strongest published attention-based detectors in the broader research; CodeTD outperforms them on 84\% of (model, language) cells across HumanEval, MBPP, BigCodeBench, and four MultiPL-E languages. The gap widens under transfer (Tab. 3): per-head methods like LapEigvals drop – 13-20 ROC-AUC pts cross-benchmark, while CodeTD remains more stable.

To sum up, CodeTD differs from TOHA \cite{bazarova2025hallucination} in five main respects:
\begin{enumerate}
    \item addressing execution-defined functional correctness;
    \item extending applications to best-of-N selection, multilingual code generation, and cross-benchmark transfer;
    \item introducing a richer family of prompt-generation features, including asymmetric $\mathrm{MTD}(P,G)$ and $\mathrm{MTD}(G,P)$, both $H_0$ and $H_1$, and diagonal self-attention statistics;
    \item explaining the importance of $H_1$ (cycles) (Appendix~\ref{app:cycles});
    \item interpreting MTD as manifold-topology divergence in value space (Appendix~\ref{app:hallucination_definition}).
\end{enumerate}

To further support our claims, we provide an additional comparison with TOHA in two versions (see Table \ref{tab:toha_comp}): (1) the final hallucination score is the average topological divergence from the selected heads (Algorithm 1 in \citet{bazarova2025hallucination}); (2) the prediction is obtained via classifier trained over the topological divergence features from selected heads (Table 9 in \citet{bazarova2025hallucination}). CodeTD outperforms both versions of TOHA.

\begin{table}[htb!]
\setlength{\tabcolsep}{2pt}
\centering
\begin{adjustbox}{width=0.99\columnwidth}
\begin{tabular}{lccc}
   \hline
   \textbf{Method} & \textbf{HE} & \textbf{MBPP} & \textbf{BCB} \\ \hline
   \multicolumn{4}{c}{StarCoder2-7B} \\ \hline
   TOHA (1) & $73.3 \pm 4.3$ & $74.7 \pm 2.2$ & $81.6 \pm 5.7$ \\
   TOHA (2) & $73.7 \pm 5.9$ &	$76.2 \pm 2.1$ & $81.3 \pm 6.2$ \\
   CodeTD (ours) & $\mathbf{82.5 \pm 2.1}$ & $\mathbf{81.4 \pm 3.6}$ & $\mathbf{83.7 \pm 4.3}$ \\ \hline
   \multicolumn{4}{c}{DeepSeek-Coder-6.7B} \\ \hline
   TOHA (1) & $76.4 \pm 3.9$ &	$67.6 \pm 4.3$ & $58.5 \pm 3.5$ \\
   TOHA (2) & $78.1 \pm 4.1$ & $70.3 \pm 7.8$ & $64.3 \pm 1.4$ \\
   CodeTD (ours) & $\mathbf{86.4 \pm 1.8}$ & $\mathbf{81.2 \pm 3.7}$ & $\mathbf{69.2 \pm 2.8}$ \\ \hline
   \multicolumn{4}{c}{Qwen2.5-Coder-7B} \\ \hline
   TOHA (1) & $72.1 \pm 3.9$ & $72.2 \pm 3.8$ & $64.6 \pm 3.6$ \\
   TOHA (2) & $77.5 \pm 4.9$ & $73.0 \pm 3.9$ & $63.7 \pm 4.8$ \\
   CodeTD (ours) & $\mathbf{81.6 \pm 4.7}$ & $\mathbf{80.9 \pm 3.5}$ & $\mathbf{70.0 \pm 1.4}$ \\ \hline
   \multicolumn{4}{c}{Magicoder-S-DS-6.7B} \\ \hline
   TOHA (1) & $66.7 \pm 7.2$ & $69.3 \pm 2.9$ & $58.1 \pm 3.3$ \\
   TOHA (2) & $70.6 \pm 8.9$ & $73.4 \pm 3.3$ & $65.1 \pm 2.6$ \\
   CodeTD (ours) & $\mathbf{79.3 \pm 3.5}$ & $\mathbf{80.7 \pm 2.5}$ & $\mathbf{68.4 \pm 1.5}$ \\ \hline
\end{tabular}
\end{adjustbox}
\caption{Comparison with TOHA \cite{bazarova2025hallucination}, ROC-AUC.}
\label{tab:toha_comp}
\end{table}

\begin{table}[]
    \centering
    \begin{tabular}{cc}
\hline
\# tokens & ROC-AUC \\
\hline
100 & 0.633 ± 0.034 \\
300 & 0.673 ± 0.012 \\ 
500& 0.656 ± 0.048 \\ 
Full generation	& 0.69 ± 0.02 \\
\hline
    \end{tabular}
    \caption{Performance of CodeTD when trained on partial
generations of different lengths.}
    \label{tab:quanta}
\end{table}

\section{On Statistical Significance for BCB}
\label{app:bcb_stat_significance}

For BCB, CodeTD significantly outperforms LapEigvals on \textbf{3 of 5} under a paired one-sided t-test across folds (Shapiro-Wilk does not reject normality, $p > 0.48$ in all cells):

\begin{itemize}
    \item DeepSeek-Coder-6.7B: $p = 0.003$
    \item Qwen2.5-Coder-7B: $p = 0.022$
    \item Magicoder-S-DS-6.7B: $p = 0.041$
\end{itemize}

CodeLlama-7B is a tie ($p = 0.225$, mean favors CodeTD; $n = 5$ folds gives limited power). The StarCoder2-7B cell ($83.7 \pm 4.3$ vs $85.3 \pm 2.4$) is the single case where LapEigvals wins. Combining the five paired tests via Fisher's method yields $p = 0.017$ in favor of CodeTD globally on BCB.

\section{On Streaming Variant of CodeTD}

CodeTD admits a streaming variant. Let $G_t$ be the tokens generated by step $t$. Because MTD is built from a \textit{threshold filtration} of the prompt-generation graph $\mathcal{G}_t = (P \sqcup G_t, E_t, W_t)$, we can compute $\text{MTD}_k(P, G_t)$ at every $t = k, 2k, 3k, \ldots$ and trigger an early-stop if hallucination probability is larger than some threshold.
Two facts make this \textit{efficient}: (i) the union-find used in $H_0$ persistence supports incremental edge insertion in near-constant amortized time, so $\text{MTD}_0$ can be updated in $O(|E_{t+k}| - |E_t|)$ per chunk; (ii) only $\sim 5\%$ of heads need to be tracked. This is a compelling extension for long generations (e.g., BCB and repository-level code).

\textbf{Experimental validation}. We generate 100, 300, or 500 tokens, and train CodeTD hallucination detector on partial generations. We evaluated this approach using the DeepSeek-Coder-6.7B model on the BigCodeBench benchmark. Table \ref{tab:quanta} shows results. The results show that the number of generated tokens can be reduced at the cost of a slight decrease in classification performance.

\section{Experiments with Prompt Masking}

We conducted experiments on syntactic vs. logical errors. On the HumanEval dataset, we masked out 50\% of tokens in the middle of prompts to force Code LLM to pay less attention to the prompt. Results are shown in Table \ref{tab:prompt_masking}, numbers differ from the main results since we used fewer trials. Here the direct causal mechanism ``less attention to the prompt $\to$ more functional errors, with a small change in the syntactic-error rate'' is established.

\begin{table}[]
\setlength{\tabcolsep}{1.8pt}
    \centering
    \begin{adjustbox}{width=0.99\columnwidth}
    \begin{tabular}{ccccc}
        \hline
        \textbf{model} &  \textbf{prompt}    & \textbf{pass@1,\%} & \textbf{synt. correct,\%} \\
        \hline
        CodeLlama-7B        & -     & 25.9 & 92.0\\
        CodeLlama-7B        & mask  & 15.0 & 90.3\\
        DS-Coder-6.7B & -     & 34.8 & 95.9\\
        DS-Coder-6.7B & mask  & 19.5 & 90.9\\
        \hline
    \end{tabular}
    \end{adjustbox}
    \caption{Performance in prompt masking experiments.}
    \label{tab:prompt_masking}
\end{table}

\section{Supervision Cost}
\label{app:supervision_cost}

In this section, we quantify the supervision cost. We provide a label-efficiency analysis in which the classifier is trained with an increasing number of labeled prompt groups, from $10\%$ to $90\%$, and the full training set. Importantly, for each train/test split, we keep the test set as before and perform subsampling for the train set at the prompt/problem level rather than at the generation level, preserving the no-leakage protocol. We report ROC-AUC as a function of the number of labeled examples. For each train/test split, we average results over multiple runs to account for subsampling stochasticity. Table \ref{tab:supervision_cost} shows the results.

For HumanEval, each $10\%$ comprises approximately $13$ problems and $325$ generations in total retained in train splits. For MBPP, $10\%$ – $40$ problems and $200$ generations. For BCB, $10\%$ – $91$ problems and $91$ generations.

With only $20\%$ of labeled prompts, CodeTD reaches:
\begin{itemize}
\item $93\%$ for StarCoder2-7B, DeepSeek-Coder-6.7B and $90\%$ for Qwen2.5-Coder-7B, Magicoder-S-DS-6.7B of its full-data ROC-AUC for HumanEval;
\item $91\%$ for StarCoder2-7B, DeepSeek-Coder-6.7B, $89\%$ for Qwen2.5-Coder-7B and $90\%$ for Magicoder-S-DS-6.7B of its full-data ROC-AUC for MBPP;
\item $79\%$ for StarCoder2-7B, $88\%$ for DeepSeek-Coder-6.7B, $91\%$ for Qwen2.5-Coder-7B, Magicoder-S-DS-6.7B of its full-data ROC-AUC for BCB.
\end{itemize}

\begin{table}[hbt!]
\centering
\begin{tabular}{lc}
    \hline
   \textbf{Method} & \textbf{HE} \\ \hline
   \multicolumn{2}{c}{Qwen2.5-Coder-32B} \\ \hline
   CodeTD (ours) & $\mathbf{85.0 \pm 2.7}$ \\
   Zero-shot prompt & $67.4 \pm 3.5$ \\ \hline
\end{tabular}
\caption{HumanEval features for larger models. See Section \ref{app:large_models} for details.}
\label{tab:detection_qwencoder32b}
\end{table}

\begin{table}[hbt!]
\centering
\begin{tabular}{lcc}
   \hline
   \textbf{Method} & \textbf{MBPP}  \\ \hline
   CodeTD (ours)\\(from DeepSeek-Coder-6.7B) & $81.2 \pm 3.7$  \\
   DeepSeek R1 (671B model) & $89.9 \pm 2.7$ \\
  CodeTD (ours) \\(from DeepSeek-Coder-6.7B)\\ + DeepSeek R1 (671B) & $\mathbf{93.2 \pm 3.4}$  \\ \hline
\end{tabular}
\caption{MBPP features for larger models. See Section \ref{app:large_models} for details.}
\label{tab:detection_deepseek_r1}
\end{table}

\begin{table}[htb!]
\centering
\setlength{\tabcolsep}{2pt}
\begin{tabular}{lcc}
   \hline
   \textbf{Model} & \textbf{HE} & \textbf{MBPP} \\ \hline
    \multicolumn{3}{c}{Qwen2.5-Coder-1.5b} \\ \hline
    Diag. Feat. & $83.2 \pm 4.3$ & $78.8 \pm 2.2$ \\
    - w/ MTD 0-dim & $\underline{84.7 \pm 3.8}$ & $\underline{81.7 \pm 1.1}$ \\
    - w/ MTD 0,1-dim & $\mathbf{85.0 \pm 3.9}$ & $\mathbf{82.5 \pm 1.0}$ \\
    \hline
    \multicolumn{3}{c}{Qwen2.5-Coder-3b} \\ \hline
    Diag. Feat. & $75.5 \pm 6.6$ & $76.5 \pm 1.6$ \\
    - w/ MTD 0-dim & $\underline{80.8 \pm 2.5}$ & $\underline{78.8 \pm 1.5}$ \\
    - w/ MTD 0,1-dim & $\mathbf{81.3 \pm 2.5}$ & $\mathbf{79.3 \pm 1.9}$ \\
    \hline
    \multicolumn{3}{c}{StarCoder2-7B} \\ \hline
    Diag. Feat. & $\mathbf{83.4 \pm 2.0}$ & $80.1 \pm 3.3$ \\
    - w/ MTD 0-dim & $\mathbf{83.4 \pm 2.9}$ & $\underline{80.4 \pm 3.3}$ \\
    - w/ MTD 0,1-dim & $\underline{82.5 \pm 2.1}$ & $\mathbf{81.4 \pm 3.6}$ \\
    \hline
   \multicolumn{3}{c}{DeepSeek-Coder-6.7B} \\ \hline
    Diag. Feat. & $84.2 \pm 1.5$ & $81.0 \pm 3.5$ \\
    - w/ MTD 0-dim & $\underline{85.7 \pm 3.3}$ & $\mathbf{81.7 \pm 2.7}$ \\
    - w/ MTD 0,1-dim & $\mathbf{86.4 \pm 1.8}$ & $\underline{81.2 \pm 3.7}$ \\
   \hline
   \multicolumn{3}{c}{Qwen2.5-Coder-7B} \\ \hline
    Diag. Feat. & $80.0 \pm 2.8$ & $76.5 \pm 4.1$ \\
    - w/ MTD 0-dim & $\underline{80.6 \pm 4.0}$ & $\underline{79.4 \pm 3.3}$ \\
    - w/ MTD 0,1-dim & $\mathbf{81.6 \pm 4.7}$ & $\mathbf{80.9 \pm 3.5}$ \\
   \hline
   \multicolumn{3}{c}{Magicoder-S-DS-6.7B} \\ \hline
    Diag. Feat. & $\underline{79.4 \pm 2.9}$ & $\underline{80.4 \pm 1.0}$ \\
    - w/ MTD 0-dim & $\mathbf{80.5 \pm 2.9}$ & $79.5 \pm 2.7$ \\
    - w/ MTD 0,1-dim & $79.3 \pm 3.5$ & $\mathbf{80.7 \pm 2.5}$ \\
   \hline
\end{tabular}
\caption{Attention feature ablation in CodeTD.}
\label{tab:detection_ablation_he_mbpp}
\end{table}

\begin{table}
\centering
\begin{tabular}{lc}
    \hline
   \textbf{Method} & \textbf{2-Shot} \\ \hline
   \multicolumn{2}{c}{DeepSeek-Coder-6.7B} \\ \hline
   Prompt Len. & $56.0 \pm 2.0$ \\
   Gen. Len. & $54.4 \pm 1.3$ \\
   Mean Log. Prob. & $61.9 \pm 1.7$ \\
   Self-Eval & $50.0 \pm 0.0$ \\
   Interrogate-LLM & $61.1 \pm 3.1$ \\ 
   CodeTD (ours) & $\mathbf{81.5 \pm 2.0}$ \\ \hline
   \multicolumn{2}{c}{Qwen2.5-Coder-7B} \\ \hline
   Prompt Len. & $53.5 \pm 2.9$ \\
   Gen. Len. & $56.9 \pm 3.3$ \\
   Mean Log. Prob. & $56.6 \pm 3.4$ \\
   Self-Eval & $68.4 \pm 1.0$ \\
   Interrogate-LLM & $56.6 \pm 1.5$ \\
   CodeTD (ours) & $\mathbf{78.1 \pm 3.1}$ \\ \hline
\end{tabular}
\caption{Code hallucination detection with attention features for MBPP dataset with increasing complexity of prompt: two-shot prompts.}
\label{tab:complex_prompt}
\end{table}

\begin{table}
\setlength{\tabcolsep}{3.5pt}
\centering
\begin{tabular}{lcc}
    \hline
    \textbf{Classifier} & \textbf{HE} & \textbf{MBPP} \\ \hline
    \multicolumn{3}{c}{StarCoder2-7B} \\ \hline
    XGBoost & $82.5 \pm 2.1$ & $81.4 \pm 3.6$ \\
    MLP & $80.9 \pm 4.7$ & $81.1 \pm 1.7$ \\
    Logistic Regression & $\underline{84.3 \pm 3.9}$ & $\mathbf{82.9 \pm 2.3}$ \\
    SVC & $\mathbf{84.9 \pm 4.0}$ & $\underline{82.3 \pm 2.6}$ \\ \hline
    \multicolumn{3}{c}{CodeLlama-7B} \\ \hline
    XGBoost & $\mathbf{85.6 \pm 3.9}$ & $\mathbf{83.4 \pm 3.3}$ \\
    MLP & $81.8 \pm	7.2$ & $81.6 \pm 2.2$ \\
    Logistic Regression & $81.8 \pm 7.1$ & $\underline{82.6 \pm 2.1}$ \\
    SVC & $\underline{83.2 \pm 5.4}$ & $82.4 \pm 1.2$ \\ \hline
    \multicolumn{3}{c}{DeepSeek-Coder-6.7B} \\ \hline
   XGBoost & $\underline{86.4 \pm 1.8}$ & $81.2 \pm 3.7$ \\ 
   MLP & $84.3 \pm 2.4$ & $\underline{81.7 \pm 1.1}$ \\ 
   Logistic Regression & $85.8 \pm 3.2$ & $\mathbf{81.8 \pm 2.4}$ \\
   SVC & $\mathbf{87.0 \pm 2.4}$ & $81.4 \pm 1.6$ \\ \hline
    \multicolumn{3}{c}{Qwen2.5-Coder-7B} \\ \hline
   XGBoost & $\mathbf{81.6 \pm 4.7}$ & $\underline{80.9 \pm 3.5}$ \\
   MLP & $\underline{81.3 \pm 1.6}$ & $79.5 \pm 2.0$ \\
   Logistic Regression & $80.0 \pm 1.6$ & $\mathbf{81.0 \pm 1.9}$ \\
   SVC & $79.6 \pm 1.7$ & $\mathbf{81.0 \pm 1.7}$ \\ \hline
    \multicolumn{3}{c}{Magicoder-S-DS-6.7B} \\ \hline
   XGBoost & $79.3 \pm 3.5$ & $80.7 \pm 2.5$ \\
   MLP & $76.5 \pm 3.3$ & $78.6 \pm 3.6$ \\
   Logistic Regression & $\mathbf{81.7 \pm 1.6}$ & $\underline{81.3 \pm 2.8}$ \\
   SVC & $\underline{80.5 \pm 2.4}$ & $\mathbf{82.5 \pm 2.6}$ \\ \hline
\end{tabular}
\caption{Ablation study for the choice of a classification model in CodeTD.}
\label{tab:he_mbpp_mlp_vs_xgboost}
\end{table}

\begin{table*}
\setlength{\tabcolsep}{1.5pt}
\centering
\begin{adjustbox}{width=\textwidth}
\begin{tabular}{m{2.5cm}|c|c|c|c|c|c|c|c|c|c}
    \hline
    \textbf{$\%$ of problems retained} & $\mathbf{10\%}$ & $\mathbf{20\%}$ & $\mathbf{30\%}$ & $\mathbf{40\%}$ & $\mathbf{50\%}$ & $\mathbf{60\%}$ & $\mathbf{70\%}$ & $\mathbf{80\%}$ & $\mathbf{90\%}$ & $\mathbf{100\%}$  \\ \hline
    \multicolumn{11}{c}{HumanEval} \\ \hline
   StarCoder2-7B & $71.0 \pm 6.0$ & $76.8 \pm 4.1$ & $78.6 \pm 4.2$ & $80.7 \pm 3.0$ & $81.2 \pm 3.0$ & $81.7 \pm 3.5$ & $81.9 \pm 2.6$ & $83.0 \pm 2.9$ & $82.7 \pm 2.4$ & $82.5 \pm 2.1$ \\
   DeepSeek-Coder-6.7B & $75.2 \pm 5.7$ & $80.6 \pm 4.8$ &	$82.0 \pm 3.0$ & $83.0 \pm 2.5$ & $83.4 \pm 2.6$ & $84.0 \pm 2.9$ & $84.1 \pm 2.4$ & $84.9 \pm 2.0$ & $85.5 \pm 1.8$ &	$86.4 \pm 1.8$ \\
   Qwen2.5-Coder-7B & $69.0 \pm 7.1$ & $74.0 \pm 5.1$ & $77.0 \pm 4.2$ & $77.2 \pm 5.3$ & $78.2 \pm 4.8$ & $79.4 \pm 4.6$ & $80.7 \pm 4.2$ & $81.2 \pm 4.0$ & $81.7 \pm 3.9$ & $81.6 \pm 4.7$ \\
   Magicoder-S-DS-6.7B & $65.6 \pm 7.2$ & $71.4 \pm 5.3$ &	$72.8 \pm 4.8$ & $75.4 \pm 4.8$ & $76.8 \pm 3.9$ & $77.0 \pm 2.7$ & $78.7 \pm 3.5$ & $78.5 \pm 3.4$ & $80.1 \pm 3.9$ &	$79.3 \pm 3.5$ \\ \hline
   \multicolumn{11}{c}{MBPP} \\ \hline
   StarCoder2-7B &	$67.9 \pm 4.7$ & $74.3 \pm 2.9$ & $76.7 \pm 2.8$ & $77.8 \pm 2.8$ & $79.0 \pm 3.3$ & $79.8 \pm 3.4$ & $79.7 \pm 3.2$ & $80.7 \pm 3.0$ & $80.8 \pm 2.8$ & $81.4 \pm 3.6$ \\
   DeepSeek-Coder-6.7B & $68.5 \pm 5.9$ & $74.6 \pm 4.2$ &	$75.5 \pm 4.4$ & $77.5 \pm 3.9$ & $78.7 \pm 3.8$ & $78.9 \pm 3.7$ & $79.7 \pm 3.2$ & $79.9 \pm 3.6$ & $80.6 \pm 3.0$ &	$81.2 \pm 3.7$ \\
   Qwen2.5-Coder-7B & $68.1 \pm 5.2$ & $72.7 \pm 4.2$ & $74.9 \pm 3.8$ & $76.4 \pm 3.4$ & $77.2 \pm 3.4$ & $78.1 \pm 4.0$ & $78.6 \pm 3.7$ & $79.7 \pm 4.3$ & $80.2 \pm 3.9$ & $80.9 \pm 3.5$ \\
   Magicoder-S-DS-6.7B & $66.6 \pm 5.9$ & $73.2 \pm 4.8$ &	$74.9 \pm 3.7$ & $76.4 \pm 3.7$ & $77.3 \pm 3.0$ & $78.5 \pm 2.7$ & $79.3 \pm 2.5$ & $80.1 \pm 2.2$ & $79.7 \pm 2.5$ &	$80.7 \pm 2.5$ \\ \hline
   \multicolumn{11}{c}{BCB} \\ \hline
   StarCoder2-7B & $62.6 \pm 8.9$ & $66.5 \pm 10.1$ & $73.5 \pm 7.2$ & $78.1 \pm 5.8$ & $80.2 \pm 4.7$ & $80.4 \pm 3.8$ & $81.2 \pm 3.8$ & $82.0 \pm 4.4$ & $81.8 \pm 3.2$ & $83.7 \pm 4.3$ \\
   DeepSeek-Coder-6.7B & $58.9 \pm 3.9$ & $61.4 \pm 4.6$ &	$64.6 \pm 2.6$ & $66.1 \pm 2.6$ & $67.6 \pm 2.8$ & $67.7 \pm 2.4$ & $68.9 \pm 3.1$ & $68.2 \pm 2.9$ & $69.3 \pm 2.5$ & $69.2 \pm 2.8$ \\
   Qwen2.5-Coder-7B & $61.4 \pm 3.9$ & $64.0 \pm 3.8$ & $63.9 \pm 3.0$ & $65.7 \pm 3.7$ & $66.4 \pm 3.2$ & $67.3 \pm 3.2$ & $68.6 \pm 2.8$ & $68.7 \pm 3.1$ & $69.7 \pm 3.5$ & $70.0 \pm 1.4$ \\
   Magicoder-S-DS-6.7B & $60.2 \pm 3.2$ & $62.6 \pm 3.7$ & $64.3 \pm 3.2$ & $65.9 \pm 2.2$ & $67.2 \pm 2.3$ & $67.0 \pm 2.6$ & $67.2 \pm 2.2$ & $67.9 \pm 2.5$ & $68.7 \pm 2.5$ & $68.4 \pm 1.5$ \\
   \hline
\end{tabular}
\end{adjustbox}
\caption{Label-efficiency analysis quantifying supervision cost of CodeTD.}
\label{tab:supervision_cost}
\end{table*}

\begin{table}[hbt!]
\centering
\begin{tabular}{lc}
    \hline
   \textbf{Method} & \textbf{HE} \\ \hline
   \multicolumn{2}{c}{CodeLlama-34B} \\ \hline
   Prompt Len. & $57.2 \pm 6.6$ \\
   Gen. Len. & $62.7 \pm 1.8$ \\
   Mean Log. Prob. & $72.9 \pm 3.4$ \\ 
   CodeT5-base ft. & $54.7 \pm 5.5$ \\
   CodeTD (ours) & $\mathbf{84.1 \pm 3.2}$ \\ \hline
   \multicolumn{2}{c}{Qwen2.5-Coder-32B} \\ \hline
   Prompt Len. & $53.2 \pm 9.2$ \\
   Gen. Len. & $58.0 \pm 3.1$ \\
   Mean Log. Prob. & $67.0 \pm 4.0$ \\ 
   CodeT5-base ft. & $53.3 \pm 4.5$ \\
   CodeTD (ours) & $\mathbf{85.0 \pm 2.7}$ \\ \hline
   \multicolumn{2}{c}{DeepSeek-Coder-33B} \\ \hline
   Prompt Len. & $53.3 \pm 6.5$ \\
   Gen. Len. & $56.5 \pm 4.2$ \\
   Mean Log. Prob. & $69.9 \pm 2.6$ \\ 
   CodeT5-base ft. & $57.6 \pm 3.9$ \\
   CodeTD (ours) & $\mathbf{88.2 \pm 3.0}$ \\ \hline
\end{tabular}
\caption{Code hallucination detection for HumanEval dataset for larger models.}
\label{tab:detection_results_he_large}
\end{table}

\begin{table}
\setlength{\tabcolsep}{3pt}
\centering
\begin{tabular}{lcc}
   \hline
   \textbf{Method} & \textbf{\small HE $\rightarrow$ MBPP} &
   \textbf{\small MBPP $\rightarrow$ HE} \\ \hline
   \multicolumn{3}{c}{Qwen2.5-Coder-3b} \\ \hline
   Diag. Feat. & $58.7$ & $61.3$ \\
   - w/ MTD 0-dim & $\underline{60.6}$ & $\underline{63.8}$ \\
   - w/ MTD 0,1-dim & $\mathbf{60.8}$ & $\mathbf{65.7}$ \\
    \hline
   \multicolumn{3}{c}{StarCoder2-7B} \\ \hline
   Diag. Feat. & $67.5$ & $\mathbf{68.3}$ \\
   - w/ MTD 0-dim & $\mathbf{71.4}$ & $\underline{67.4}$ \\ 
   - w/ MTD 0,1-dim & $\underline{67.7}$ & $66.0$ \\
   \hline
   \multicolumn{3}{c}{DeepSeek-Coder-6.7B} \\ \hline
   Diag. Feat. & $\underline{65.5}$ & $63.5$ \\
   - w/ MTD 0-dim & $64.5$ & $\underline{69.8}$ \\ 
   - w/ MTD 0,1-dim & $\mathbf{69.9}$ & $\mathbf{72.4}$ \\
   \hline
   \multicolumn{3}{c}{Qwen2.5-Coder-7B} \\ \hline
   Diag. Feat. & $64.6$ & $\underline{62.9}$ \\
   - w/ MTD 0-dim. & $\underline{70.3}$ & $60.7$ \\
   - w/ MTD 0,1-dim & $\mathbf{70.9}$ & $\mathbf{65.8}$ \\
   \hline
\end{tabular}
\caption{Dataset transferability attention feature ablation.}
\label{tab:transfer_ablation}
\end{table}

\section{Licenses}

Licenses of pretrained models and benchmarks used in this paper permit use for research purposes.

\begin{table}[hbt!]
\setlength{\tabcolsep}{4pt}
\centering
\begin{tabular}{lcc}
    \hline
   \textbf{Method} & \textbf{Greedy Dec.} & \textbf{T=0.8} \\ \hline
   \multicolumn{3}{c}{StarCoder2-7B} \\ \hline
   CodeJudge w/o ref. & $70.6 \pm 4.7$ & $69.6 \pm 1.9$ \\
   CodeJudge w/ ref. & $\underline{75.8 \pm 4.7}$ & $\underline{71.1 \pm 2.8}$ \\
   CodeTD (ours) & $\mathbf{82.2 \pm 4.3}$ & $\mathbf{82.5 \pm 2.1}$ \\ \hline
   \multicolumn{3}{c}{CodeLlama-7B} \\ \hline
   CodeJudge w/o ref. & $71.3 \pm 4.1$ & $\underline{72.1 \pm 1.3}$ \\
   CodeJudge w/ ref. & $\underline{73.0 \pm 3.3}$ & $71.9 \pm 3.6$ \\
   CodeTD (ours) & $\mathbf{78.1 \pm 4.3}$ & $\mathbf{85.6 \pm 3.9}$ \\ \hline
   \multicolumn{3}{c}{DeepSeek-Coder-6.7B} \\ \hline
   CodeJudge w/o ref. & $\underline{69.2 \pm 3.2}$ & $\underline{68.4 \pm 1.1}$ \\
   CodeJudge w/ ref. & $68.8 \pm 7.9$ & $68.3 \pm 3.1$ \\
   CodeTD (ours) & $\mathbf{83.9 \pm 8.4}$ & $\mathbf{86.4 \pm 1.8}$ \\ \hline
   \multicolumn{3}{c}{Qwen2.5-Coder-7B} \\ \hline
   CodeJudge w/o ref. & $67.2 \pm 5.7$ & $69.1 \pm 3.0$ \\
   CodeJudge w/ ref. & $\underline{70.4 \pm 7.4}$ & $\underline{69.3 \pm 2.6}$ \\
   CodeTD (ours) & $\mathbf{73.7 \pm 5.2}$ & $\mathbf{81.6 \pm 4.7}$ \\ \hline
   \multicolumn{3}{c}{Magicoder-S-DS-6.7B}\\ \hline
   CodeJudge w/o ref. & $55.3 \pm 2.7$ & $60.4 \pm 1.8$ \\
   CodeJudge w/ ref. & $\underline{57.9 \pm 7.1}$ & $\underline{61.2 \pm 2.9}$ \\
   CodeTD (ours) & $\mathbf{68.3 \pm 3.5}$ & $\mathbf{79.3 \pm 3.5}$ \\ \hline
    \multicolumn{3}{c}{CodeLlama-34B} \\ \hline
    CodeJudge w/o ref. & $65.3 \pm 3.0$ & $70.7 \pm 5.1$ \\
    CodeJudge w/ ref. & $\underline{68.7 \pm 3.8}$ & $\underline{73.2 \pm 4.4}$ \\
    CodeTD (ours) & $\mathbf{75.0 \pm 5.1}$ & $\mathbf{84.1 \pm 3.2}$ \\ \hline
    \multicolumn{3}{c}{DeepSeek-Coder-33B} \\ \hline
    CodeJudge w/o ref. & $72.0 \pm 7.2$ & $69.0 \pm 2.8$ \\
    CodeJudge w/ ref. & $\underline{75.8 \pm 5.6}$ & $\underline{70.9 \pm 1.3}$ \\
    CodeTD (ours) & $\mathbf{86.5 \pm 7.2}$ & $\mathbf{88.2 \pm 3.0}$ \\ \hline
    \multicolumn{3}{c}{Qwen2.5-Coder-32B} \\ \hline
    CodeJudge w/o ref. & $61.6 \pm 7.3$ & $66.5 \pm 3.5$ \\
    CodeJudge w/ ref. & $\underline{62.0 \pm 7.5}$ & $\underline{68.2 \pm 3.7}$ \\
    CodeTD (ours) & $\mathbf{74.0 \pm 6.4}$ & $\mathbf{85.0 \pm 2.7}$ \\ \hline
\end{tabular}
\caption{Comparison with CodeJudge A. S. \cite{tong2024codejudge} on HumanEval in two setups: code generation with greedy decoding and sampling with T=0.8.}
\label{tab:codejudge}
\end{table}

\begin{table}
\centering
\begin{adjustbox}{height=0.485\textheight}
\begin{tabular}{lcc}
   \hline
   \textbf{Method} & \textbf{HE} & \textbf{MBPP} \\ \hline
   \multicolumn{3}{c}{StarCoder2-7B} \\ \hline
   Prompt. Len. & $68.9 \pm 8.7$ & $52.0 \pm 8.1$ \\
   Gen. Len. & $55.6 \pm 7.2$ & $49.4 \pm 2.4$ \\
   Mean Log. Prob. & $62.0 \pm 10.7$ & $57.1 \pm 8.1$ \\
   Pylint & $56.7 \pm 2.6$ & $54.5 \pm 0.9$ \\
   CodeT5-base ft. & $60.2 \pm 4.8$ & $56.5 \pm 5.4$ \\
   Self-Eval & $52.0 \pm 2.6$ & $61.2 \pm 4.8$ \\
   AttnLogDet & $70.6 \pm 7.5$ & $72.8 \pm 2.8$ \\
   AttnEigvals & $73.6 \pm 7.5$ & $62.5 \pm 6.3$ \\
   LapEigvals & $\underline{76.4 \pm 4.4}$ & $\underline{74.5 \pm 4.5}$ \\
   CodeTD (ours) & $\mathbf{82.2 \pm 4.3}$ & $\mathbf{79.2 \pm 6.1}$ \\ \hline
   \multicolumn{3}{c}{CodeLlama-7B} \\ \hline
   Prompt. Len. & $53.2 \pm 10.3$ & $61.8 \pm 4.2$ \\
   Gen. Len. & $59.9 \pm 5.6$ & $54.6 \pm 5.7$ \\
   Mean Log. Prob. & $56.6 \pm 14.0$ & $64.5 \pm 6.2$ \\
   Pylint & $55.3 \pm 3.0$ & $53.4 \pm 1.6$ \\
   CodeT5-base ft. & $64.1 \pm 7.1$ & $54.7 \pm 8.0$ \\
   Self-Eval & $47.9 \pm 3.7$ & $50.0 \pm 0.0$ \\
   AttnLogDet & $74.5 \pm 7.7$ & $72.7 \pm 3.9$ \\
   AttnEigvals & $\mathbf{81.1 \pm 6.2}$ & $69.0 \pm 7.5$ \\
   LapEigvals & $73.3 \pm 7.7$ & $\underline{77.0 \pm 4.2}$ \\
   CodeTD (ours) & $\underline{78.1 \pm 4.3}$ & $\mathbf{82.8 \pm 3.2}$ \\ \hline
   \multicolumn{3}{c}{DeepSeek-Coder-6.7B} \\ \hline
   Prompt. Len. & $59.1 \pm 6.6$ & $52.0 \pm 6.2$ \\
   Gen. Len. & $45.5 \pm 7.9$ & $55.2 \pm 5.4$ \\
   Mean Log. Prob. & $64.1 \pm 6.6$ & $63.5 \pm 4.3$ \\
   Pylint & $52.3 \pm 2.0$ & $52.6 \pm 2.7$ \\
   CodeT5-base ft. & $58.4 \pm 10.3$ & $49.4 \pm 3.1$ \\
   Self-Eval & $52.9 \pm 6.4$ & $50.0 \pm 0.0$ \\
   AttnLogDet & $\underline{80.3 \pm 6.0}$ & $76.4 \pm 3.7$ \\
   AttnEigvals & $79.5 \pm 8.7$ & $72.9 \pm 4.6$ \\
   LapEigvals & $79.5 \pm 5.7$ & $\underline{77.2 \pm 2.3}$ \\
   CodeTD (ours) & $\mathbf{83.9 \pm 8.4}$ & $\mathbf{79.3 \pm 2.4}$ \\ \hline
   \multicolumn{3}{c}{Qwen2.5-Coder-7B} \\ \hline
   Prompt. Len. & $53.6 \pm 8.7$ & $57.5 \pm 7.2$ \\
   Gen. Len. & $54.0 \pm 11.0$ & $54.6 \pm 3.8$ \\
   Mean Log. Prob. & $54.6 \pm 8.7$ & $61.7 \pm 6.6$ \\
   Pylint & $64.0 \pm 3.0$ & $61.6 \pm 4.0$ \\
   CodeT5-base ft. & $59.0 \pm 7.3$ & $54.1 \pm 4.2$ \\
   Self-Eval & $\underline{72.1 \pm 7.8}$ & $65.7 \pm 5.0$ \\
   AttnLogDet & $61.9 \pm 8.3$ & $68.9 \pm 4.2$ \\
   AttnEigvals & $63.7 \pm 8.2$ & $65.7 \pm 5.1$ \\
   LapEigvals & $65.9 \pm 12.3$ & $\underline{72.1 \pm 4.1}$ \\
   CodeTD (ours) & $\mathbf{73.7 \pm 5.2}$ & $\mathbf{74.0 \pm 2.7}$ \\ \hline
   \multicolumn{3}{c}{Magicoder-S-DS-6.7B} \\ \hline
   Prompt. Len. & $53.0 \pm 6.1$ & $58.3 \pm 4.1$ \\
   Gen. Len. & $59.0 \pm 4.8$ & $52.1 \pm 2.3$ \\
   Mean Log. Prob. & $61.4 \pm 4.8$ & $61.1 \pm 4.3$ \\
   Pylint & $50.8 \pm 1.5$ & $50.8 \pm 0.7$ \\
   CodeT5-base ft. & $57.8 \pm 15.3$ & $54.5 \pm 7.7$ \\
   Self-Eval & $57.3 \pm 8.6$ & $49.1 \pm 2.4$ \\
   AttnLogDet & $64.5 \pm 8.5$ & $75.7 \pm 2.9$ \\
   AttnEigvals & $63.1 \pm 8.4$ & $72.7 \pm 5.4$ \\
   LapEigvals & $\underline{66.9 \pm 6.2}$ & $\underline{78.1 \pm 3.6}$ \\
   CodeTD (ours) & $\mathbf{68.3 \pm 3.5}$ & $\mathbf{80.6 \pm 1.9}$ \\ \hline
\end{tabular}
\end{adjustbox}
\caption{ROC-AUC of code hallucination detection for generation with greedy decoding.}
\label{tab:detection_results_greedy}
\end{table}

\begin{table*}
    \centering
    \begin{tabular}{lcccccc}
    \hline
         \textbf{Feature} & \multicolumn{5}{c}{\textbf{HE}} & \textbf{MBPP} \\
         \hline 
         & \textbf{Python} & \textbf{Go} & \textbf{Rust} & \textbf{Java} & \textbf{Lua} & \textbf{Python} \\ 
         \hline
    \multicolumn{6}{c}{StarCoder2-7B} \\ \hline
    avg. prompt's self-attention, layer 14, head 0 & $70.8$ & $76.8$ & $74.6$ & $68.9$ & $70.3$ & $55.2$ \\
    avg. prompt's self-attention, layer 15, head 5 & $71.1$ & $78.4$ & $75.1$ & $72.9$ & $72.4$ & $58.1$ \\
    avg. prompt's self-attention, layer 23, head 20 & $69.2$ & $72.2$ & $74.2$ & $64.7$ & $67.7$ & $50.7$ \\ \hline

    \multicolumn{6}{c}{CodeLlama-7B} \\ \hline
-MTD$_1$(P, G)/\textbar{P}\textbar, layer 15, head 27 & 67.6 & 71.1 & 73.7 & 71.0 & -- & 66.1 \\
avg. prompt's self-attention, layer 7, head 22 & 74.6 & 75.7 & 76.5 & 69.7 & -- & 63.1 \\
MTD$_0$(P, G)/\textbar{P}\textbar, layer 11, head 23 & 69.1 & 71.0 & 71.9 & 65.4 & -- & 69.1 \\ \hline
    
    \multicolumn{6}{c}{DeepSeek-Coder-6.7B} \\ \hline
    MTD$_0$(P, G)/\textbar{P}\textbar, layer 30, head 11 & $67.5$ & $65.7$ & $67.0$ & $66.3$ & $69.1$ & $58.1$ \\
    avg. prompt's self-attention, layer 12, head 17 & $65.0$ & $73.6$ & $70.0$ & $69.8$ & $69.8$ & $58.2$ \\
    avg. prompt's self-attention, layer 12, head 23 & $64.1$ & $71.0$ & $66.8$ & $64.4$ & $67.6$ & $58.2$ \\ \hline
    
    \multicolumn{6}{c}{Qwen2.5-Coder-7B} \\ \hline
    avg. generation's self-attention, layer 24, head 2 & $65.8$ & $60.2$ & $58.8$ & $60.2$ & $66.7$ & $60.8$ \\
    MTD$_0$(P, G)/\textbar{P}\textbar, layer 11, head 5 & $66.2$ & $68.8$ & $59.5$ & $61.6$ & $72.7$ & $59.4$ \\
    avg. prompt's self-attention, layer 9, head 26 & $65.4$ & $75.8$ & $66.9$ & $67.0$ & $71.1$ & $61.0$ \\ \hline
    
    \multicolumn{6}{c}{Magicoder-S-DS-6.7B} \\ \hline
    MTD$_0$(P, G)/\textbar{P}\textbar, layer 30, head 11 & $68.8$ & $70.2$ & $60.0$ & $65.1$ & $63.8$ & $58.2$ \\
    avg. prompt's self-attention, layer 13, head 13 & $63.1$ & $64.9$ & $69.0$ & $67.3$ & $63.2$ & $58.8$ \\
    avg. prompt's self-attention, layer 12, head 17 & $63.0$ & $63.2$ & $68.0$ & $67.1$ & $60.8$ & $58.9$ \\ \hline

    \end{tabular}
\caption{Top-performing features across several programming languages and benchmarks.}
\label{tab:more_languages}
\end{table*}

\begin{table*}
\begin{adjustbox}{width=\textwidth}
\setlength{\tabcolsep}{4.5pt}
\centering
\begin{tabular}{l|cc|cc|cccc|c}
    \hline
    \multirow{3}{*}{\textbf{Model}} & \multicolumn{2}{c}{\textbf{HumanEval}} & \multicolumn{2}{c}{\textbf{MBPP}} & \multicolumn{4}{c}{\textbf{MultiPL-E}} & \textbf{BCB} \\
   & \multicolumn{2}{c}{Python} & \multicolumn{2}{c}{Python} & Java & Go & Rust & Lua & Python \\
   & GD & T = 0.8 & GD & T = 0.8 & T = 0.8 & T = 0.8 & T = 0.8 & T = 0.8 & T = 0.8 \\ \hline
   StarCoder2-7B & $35.4$ & $28.9$ & $46.2$ & $42.8$ & $24.5$ & $17.5$ & $20.9$ & $19.1$ & $5.9$ \\
   CodeLlama-7B & $29.3$ & $25.9$ & $38.8$ & $35.2$ & $25.8$ & $17.6$ & $20.8$ & $0.0$ & $21.5$ \\
   DeepSeek-Coder-6.7B & $48.8$ & $40.3$ & $59.2$ & $52.6$ & $33.5$ & $23.6$ & $28.7$ & $16.6$ & $32.5$ \\
   Qwen2.5-Coder-7B & $57.3$ & $47.8$ & $58.6$ & $52.1$ & $22.7$ & $11.9$ & $22.6$ & $23.5$ & $36.5$ \\ 
   Magicoder-S-DS-6.7B & $73.2$ & $65.5$ & $63.2$ & $61.3$ & $48.9$ & $40.3$ & $44.2$ & $34.6$ & $36.6$ \\ \hline
\end{tabular}
\end{adjustbox}
\caption{Characteristics of generated data, pass@1. We use two regimes to generate solutions -- greedy decoding (GD) and sampling with temperature $T = 0.8$.}
\label{tab:gen_data_stat_total}
\end{table*}

\begin{table*}
\centering
\begin{adjustbox}{width=\textwidth}
\begin{tabular}{lcccc|c}
    \hline
    & \rotatebox{45}{\textbf{StarCoder2-7B}} & \rotatebox{45}{\textbf{DeepSeek-Coder-6.7B}} & \rotatebox{45}{\textbf{Qwen2.5-Coder-7B}} & \rotatebox{45}{\textbf{Magicoder-S-DS-6.7B}} & \rotatebox{45}{\textbf{DeepSeek-Coder-33B}} \\ \hline
    StarCoder2-7B & $82.5 \pm 2.1$ & $89.5 \pm 2.7$ & $87.0 \pm 2.9$ & $\mathbf{91.7 \pm 2.1}$ & $\underline{90.3 \pm 1.1}$ \\ 
    DeepSeek-Coder-6.7B & $81.0 \pm 1.7$ & $86.4 \pm 1.8$ & $85.7 \pm 1.9$ & $\mathbf{89.0 \pm 1.6}$ & $\underline{87.9 \pm 2.3}$ \\ 
    Qwen2.5-Coder-7B & $80.4 \pm 2.5$ & $85.3 \pm 2.0$ & $81.6 \pm 4.7$ & $\mathbf{87.7 \pm 1.6}$ & $\underline{86.3 \pm 3.1}$ \\ 
    Magicoder-S-DS-6.7B & $70.9 \pm 3.4$ & $74.7 \pm 3.5$ & $79.0 \pm 1.3$ & $\underline{79.3 \pm 3.5}$ & $\mathbf{81.2 \pm 4.1}$ \\ 
    \hline
\end{tabular}
\end{adjustbox}
\caption{Cross-model transferability: rows correspond to code-generating models, columns correspond to feature-extracting models.}
\label{tab:cross_model}
\end{table*}

\begin{table*}
\centering
\begin{adjustbox}{width=\textwidth}
\begin{tabular}{lccc|ccc}
    \hline
    \textbf{Model} & \multicolumn{3}{c}{\textbf{HE}} & \multicolumn{3}{c}{\textbf{MBPP}} \\ \hline 
    & \textbf{Top-1} & \textbf{Top-2} & \textbf{Top-3} & \textbf{Top-1} & \textbf{Top-2} & \textbf{Top-3} \\ \hline
   StarCoder2-7B & P $17.2 \pm 4.5$ &	AE $2.9 \pm 2.2$ & NE $0.8 \pm 0.5$ & P $15.0 \pm 3.5$ & AE $9.7 \pm 2.2$ & NE $0.7 \pm 0.4$ \\
   CodeLlama-7B & P $14.6 \pm 3.6$ & AE $3.8 \pm 2.7$ & NE $0.2 \pm 0.2$ & P $15.5 \pm 1.6$ & AE $7.4 \pm 1.2$ & NE $0.5 \pm 0.4$ \\
   DeepSeek-Coder-6.7B & P $14.6 \pm 2.9$ & AE $4.9 \pm 1.2$ &	IE $0.5 \pm 0.6$ & AE $12.6 \pm 3.2$ & P $11.1 \pm 3.8$ & NE $0.7 \pm 0.5$ \\
   Qwen2.5-Coder-7B & P $16.0 \pm 5.4$ & AE $6.9 \pm 2.8$ & NE $2.1 \pm 0.7$ & P $11.3 \pm 0.5$ & AE $10.8 \pm 1.4$ & NE $1.3 \pm 0.5$ \\
   Magicoder-6.7B & AE $12.1 \pm 4.8$ & P $10.4 \pm 4.8$ & IE $0.7 \pm 0.6$ & AE $14.5 \pm 4.4$ & P $9.2 \pm 2.3$ & TE $0.6 \pm 0.5$ \\ \hline
\end{tabular}
\end{adjustbox}
\caption{Failure case study. ``P'' - Passed, ``AE'' - AssertionError, ``NE'' - NameError, ``IE'' - IndexError, ``TE'' - TypeError.}
\label{tab:failure_case}
\end{table*}

\begin{table*}[]
\setlength{\tabcolsep}{3.5pt}
    \centering
    \begin{tabular}{lcccc}
    \hline
    \bf{Code LLM } & \multicolumn{2}{c}{\bf{HumanEval}} &   \multicolumn{2}{c}{\bf{MBPP}} \\
             & Assertion error, \% & Other errors, \% & Assertion error, \% & Other errors, \% 
                \\
        \hline
        StarCoder2-7B  & 71.5 & 28.5 & 85.1 & 14.9 \\
        CodeLlama-7B & 79.3 & 20.7 & 83.4 & 16.5 \\
        DeepSeek-Coder-6.7B-base & 81.7 & 18.3 & 86.3 & 13.4 \\
        Qwen2.5-Coder-7B & 62.9 & 37.1 & 66.9 & 33.1 \\ 
        Magicoder-S-DS-6.7B & 83.1 & 16.9 & 90.6 & 9.4\\
        \hline
    \end{tabular}
    \caption{Breakdown of Code LLMs by error type.}
    \label{tab:assertion_error_types}
\end{table*}

\begin{table*}
\setlength{\tabcolsep}{2pt}
\centering
\begin{tabular}{lcccc}
   \hline
   \textbf{Method} & \textbf{Java} & \textbf{Go} & \textbf{Rust} & \textbf{Lua}  \\ \hline
   \multicolumn{5}{c}{StarCoder2-7B} \\ \hline
   AttnLogDet & $80.7 \pm 6.2$ & $82.4 \pm 4.8$ & $\underline{77.8 \pm 5.9}$ & $74.6 \pm 3.0$ \\
   AttnEigvals & $\underline{81.6 \pm 6.2}$ & $80.4 \pm 4.1$ & $75.2 \pm 8.5$ & $77.8 \pm 7.8$ \\
   LapEigvals & $\underline{81.6 \pm 5.4}$ & $\underline{82.5 \pm 3.3}$ & $77.1 \pm 5.3$ & $\underline{78.8 \pm 5.1}$ \\
   CodeTD (ours) & $\mathbf{82.5 \pm 5.6}$ & $\mathbf{86.6 \pm 4.6}$ & $\mathbf{82.5 \pm 5.8}$ & $\mathbf{82.0 \pm 2.8}$ \\ \hline
   \multicolumn{5}{c}{CodeLlama-7B} \\ \hline
   AttnLogDet & $\mathbf{85.2 \pm 4.4}$ & $55.5 \pm 10.4$ & $63.4 \pm 10.3$ & -- \\
   AttnEigvals & $\underline{78.2 \pm 9.6}$ & $66.7 \pm 10.7$ & $\underline{76.7 \pm 12.8}$ & -- \\
   LapEigvals & $75.7 \pm 8.7$ & $\underline{68.6 \pm 16.6}$ & $74.3 \pm 9.2$ & -- \\
   CodeTD (ours) & $76.8 \pm 6.1$ & $\mathbf{81.9 \pm 6.2}$ & $\mathbf{77.5 \pm 10.8}$ & -- \\ \hline
   \multicolumn{5}{c}{DeepSeek-Coder-6.7B} \\ \hline
   AttnLogDet & $80.1 \pm 5.5$ & $81.9 \pm 3.4$ & $78.3 \pm 4.2$ & $74.7 \pm 7.8$ \\
   AttnEigvals & $78.2 \pm 5.8$ & $79.0 \pm 5.2$ & $74.5 \pm 4.4$ & $84.6 \pm 3.8$ \\
   LapEigvals & $\underline{83.1 \pm 4.5}$ & $\mathbf{89.0 \pm 2.5}$ & $\underline{82.1 \pm 5.8}$ & $\underline{86.1 \pm 2.7}$ \\
   CodeTD (ours) & $\mathbf{84.5 \pm 4.8}$ & $\underline{85.2 \pm 3.1}$ & $\mathbf{82.3 \pm 7.8}$ & $\mathbf{86.2 \pm 4.7}$ \\ \hline
   \multicolumn{5}{c}{Qwen2.5-Coder-7B} \\ \hline
   AttnLogDet & $65.7 \pm 9.1$ & $76.7 \pm 4.5$ & $72.4 \pm 5.3$ & $80.3 \pm 4.0$ \\
   AttnEigvals & $78.3 \pm 2.9$ & $85.9 \pm 4.0$ & $\underline{84.0 \pm 3.2}$ & $85.8 \pm 2.5$ \\
   LapEigvals & $\underline{80.8 \pm 3.7}$ & $\underline{87.9 \pm 2.8}$ & $79.3 \pm 5.4$ & $\underline{87.3 \pm 2.4}$ \\
   CodeTD (ours) & $\mathbf{81.7 \pm 4.6}$ & $\mathbf{91.1 \pm 1.7}$ & $\mathbf{87.5 \pm 2.3}$ & $\mathbf{90.1 \pm 3.2}$ \\ \hline
   \multicolumn{5}{c}{Magicoder-S-DS-6.7B} \\ \hline
   AttnLogDet & $74.5 \pm 3.0$ & $70.2 \pm 4.7$ & $69.8 \pm 3.9$ & $68.8 \pm 3.7$ \\
   AttnEigvals & $69.2 \pm 5.8$ & $70.9 \pm 2.9$ & $\underline{69.9 \pm 6.9}$ & $75.4 \pm 2.7$ \\
   LapEigvals & $\underline{76.2 \pm 1.5}$ & $\mathbf{84.0 \pm 2.6}$ & $\mathbf{74.8 \pm 5.3}$ & $\underline{79.6 \pm 0.3}$ \\
   CodeTD (ours) & $\mathbf{77.8 \pm 4.5}$ & $\underline{80.7 \pm 1.9}$ & $\mathbf{74.8 \pm 6.1}$ & $\mathbf{80.1 \pm 2.1}$ \\ \hline
\end{tabular}
\caption{ROC-AUC for different programming languages from MultiPL-E dataset: mean and standard deviation over 5 folds.}
\label{tab:more_languages_detect_rocauc_full}
\end{table*}

\onecolumn

\section{Code Samples}
\label{app:code_samples}

\begin{verbatim}
Generation 1, Problem #53, Correct

def add(x: int, y: int):
    """Add two numbers x and y
    >>> add(2, 3)
    5
    >>> add(5, 7)
    12
    """
    return x + y

Generation 2, Problem #23, Correct

def strlen(string: str) -> int:
    """ Return length of given string
    >>> strlen('')
    0
    >>> strlen('abc')
    3
    """
    return len(string)

Generation 3, Problem #41, Not correct - hallucination
    
def car_race_collision(n: int):
    """
    Imagine a road that's a perfectly straight infinitely long line.
    n cars are driving left to right;  simultaneously, a different set of n cars
    are driving right to left.   The two sets of cars start out being very far from
    each other.  All cars move in the same speed.  Two cars are said to collide
    when a car that's moving left to right hits a car that's moving right to left.
    However, the cars are infinitely sturdy and strong; as a result, they continue moving
    in their trajectory as if they did not collide.

    This function outputs the number of such collisions.
    """
    pass
	
Generation 4, Problem #41, Not correct - hallucination

def car_race_collision(n: int):
    """
    Imagine a road that's a perfectly straight infinitely long line.
    n cars are driving left to right;  simultaneously, a different set of n cars
    are driving right to left.   The two sets of cars start out being very far from
    each other.  All cars move in the same speed.  Two cars are said to collide
    when a car that's moving left to right hits a car that's moving right to left.
    However, the cars are infinitely sturdy and strong; as a result, they continue moving
    in their trajectory as if they did not collide.

    This function outputs the number of such collisions.
    """
    # The cars are moving at the same speed. So, the distance between them is always n.
    # If we find the sum of the distances between two cars in the initial state
    # and at the end of the race, we can simply subtract them to get the number
    # of collisions.

    total_distance_before = 0
    total_distance_after = 0

    for car_distance in range(1, n + 1):
        total_distance_before += car_distance
        total_distance_after += n - car_distance + 1

    return total_distance_before - total_distance_after
\end{verbatim}

\end{document}